\documentclass[aps,pra,twocolumn,groupedaddress,superscriptaddress,bibnotes,citeautoscript,a4paper]{revtex4-2}
\usepackage{latexsym}
\usepackage{amsmath}
\usepackage{amssymb}
\usepackage{amsfonts}
\usepackage{graphicx}
\usepackage{epstopdf}
\usepackage[T1]{fontenc}
\usepackage[open]{bookmark}
\usepackage{hyperref}
\hypersetup{colorlinks=true,allcolors=blue}
\usepackage{orcidlink}
\usepackage{epsfig}
\usepackage{epstopdf}
\usepackage{soul}
\usepackage{float}
\usepackage{ragged2e}
\usepackage{color}
\usepackage{dsfont}
\usepackage[percent]{overpic}
\usepackage{lineno}
\usepackage[normalem]{ulem}
\usepackage{mathrsfs}

\begin{document}

\title{Probing strong coupling in core-shell nanoparticles with fast electron beams}

\author{Annika Brandt}
\affiliation{Institute of Theoretical Solid State Physics, Karlsruhe Institute of Technology, Kaiserstr. 12, DE-76131, Germany}

\author{Christos Tserkezis\,\orcidlink{0000-0002-2075-9036}}%
\email{ct@mci.sdu.dk}
\affiliation{POLIMA---Center for Polariton-driven Light-Matter Interactions, University of Southern Denmark, 5230 Odense M, Denmark}

\author{Carsten Rockstuhl\,\orcidlink{0000-0002-5868-0526}}%
\affiliation{Institute of Nanotechnology, Karlsruhe Institute of Technology, Kaiserstr. 12, DE-76131, Germany}
\affiliation{Institute of Theoretical Solid State Physics, Karlsruhe Institute of Technology, Kaiserstr. 12, DE-76131, Germany}
\affiliation{Center for Integrated Quantum Science and Technology, Karlsruhe Institute of Technology, Wolfgang-Gaede-Str. 1, DE-76131, Germany}

\author{P. Elli Stamatopoulou\,\orcidlink{0000-0001-9121-911X}}%
\email{elli.stamatopoulou@kit.edu}
\affiliation{Institute of Nanotechnology, Karlsruhe Institute of Technology, Kaiserstr. 12, DE-76131, Germany}
\affiliation{POLIMA---Center for Polariton-driven Light-Matter Interactions, University of Southern Denmark, 5230 Odense M, Denmark}

\date{\today}

\begin{abstract}
Collective optical excitations, such as localized surface plasmons in metallic nanoparticles and Mie resonances in high-index dielectrics, play a central role in nanoscale light–matter interactions. When such optical modes interact with electronic transitions in matter under suitable conditions, they can couple strongly, analogous to two coupled harmonic oscillators, forming hybrid light-matter states. In this work, we probe this coupling in core-shell nanoparticles using fast electrons in electron energy-loss (EEL) and cathodoluminescence (CL) spectroscopy. Owing to their highly localized fields, fast electrons can excite modes inaccessible with light-based spectroscopies, including higher-order nonradiative modes, which offer greater field confinement and potentially stronger coupling. Here, we develop an analytical framework to calculate the EEL and CL probabilities for spherical core–shell nanoparticles under aloof and penetrating electron trajectories. This formalism is applied to two representative systems: an excitonic core with a metallic shell, and a silicon core with an excitonic shell. Our main focus is to examine how the electron beam position and velocity affect our ability to probe this coupling. Depending on the electron beam parameters, we find that the spectral signature of strong coupling remains robust in plasmonic nanospheres. In contrast, it can be significantly suppressed or even completely obscured in dielectric nanospheres. Our developed formalism enables a deeper understanding of the coupling mechanisms in electron--light--matter interactions, thereby accelerating progress in single-nanoparticle-based polaritonic studies.
\end{abstract}

\maketitle

\section{Introduction}

Electron microscopy is rapidly claiming an ever more important role in the investigation of light--matter interactions at the nanoscale~\cite{abajo_rmp82,polman_natmat18}, on par with traditional~\cite{Novotny_Cambridge2006} or more advanced techniques in optical microscopy~\cite{wang_jpcm31,gao_anchem93,pohl_jmicrosc152,chen_admat31}. The unprecedented spatial and energy resolution that can nowadays be achieved in electron-beam spectroscopies~\cite{egerton_umicros107,coenen_apl99,hage_sc367,krivanek_um203}, combined with the flexibility in selectively exciting and probing optical modes that a localized source such as a swift electron can provide~\cite{vesseur_nl7,feist_umicrosc176}, opens exciting opportunities for designing and understanding novel nanoscale light--matter coupling templates. Electron energy-loss spectroscopy (EELS) played a pivotal role in the birth and growth of the field of plasmonics~\cite{ritchie_pr106,echenique_prb35} by providing the most efficient way to overcome momentum mismatch and excite surface plasmons in thin metallic films. At the same time, it has enabled the study of dark (to visible light) modes in plasmonic nanostructures~\cite{koh_nn10,chu_nl9}, and of quantum effects that become relevant in extremely small structures and tight gaps~\cite{raza_natcom6,scholl_nat483}. The complementary technique of cathodoluminescence (CL) spectroscopy further enabled the direct visualization of the local density of states (LDOS) in photonic architectures~\cite{kuttge_prb79,kociak_csr43} and the distinction between its radiative and nonradiative parts~\cite{losquin_acsphot2}, thus facilitating the control of the properties of quantum emitters at the nanoscale and accelerating the advancement of quantum nanophotonics~\cite{bozhevolnyi_nanoph6}.

Inspired by more established activities in quantum optics, quantum nanophotonics seeks to exploit the tremendous field enhancement and confinement that nanostructured optical cavities can provide, to accelerate light--matter interactions, and enter the strong-coupling regime, where quantum emitters reversibly and coherently exchange energy with the cavity for several cycles, before radiation and Ohmic losses lead to decoherence~\cite{Torma_2015,tserkezis_rpp83}. Nanophotonic environments can be typically seen as open cavities that compensate for the radiative losses from which they suffer through increased coupling strength and minimized mode volume~\cite{chikkaraddy_nat535,stuhrenberg_nl18,vasa_acsphot5}. They are thus being explored in polaritonic studies, both in fundamental research~\cite{vidal_sci373,Tserkezis2018,ciraci_nanoph8,bundgaard_josab41} and for applications in quantum technologies, e.g., for polariton lasing~\cite{kenacohen_natphot4,hakala_nc8}, transistors~\cite{ballarini_natcom4},  catalysis~\cite{xiang_chemrev124}, or cavity-modified  chemistry~\cite{ribeiro_chemsci9}. While most of these studies rely on plane-wave (laser) excitation, increasing attention is being focused on electron-beam excitation as an efficient probe of polaritonic states~\cite{wei_nl15,yankovich_nl19,crai_acsphot7,Zouros2020}.

In the quest for a clearer interpretation of the complex interaction that involves electrons, quantum emitters (e.g., in the form of excitonic transitions in semiconductors), and optical modes, (semi)analytic solutions that provide insight into the relative contribution of individual mechanisms in a computationally efficient manner are highly desirable~\cite{abajo_rmp82,Stamatopoulou:25}. Similarly to what has been done with scattering and/or transfer matrices for plane waves incident on layered structures, and Mie theory for spherical scatterers~\cite{Hohenester_Springer2020}, introducing such a formalism for electron beams can be a somewhat cumbersome, albeit fruitful endeavor~\cite{stamatopoulou_arxiv}. Solutions for electron beams interacting with planar or spherical structures have existed already since the 1950s---initially in the quasistatic approximation, where only scalar potentials are required~\cite{ritchie_pr106,echenique_prb35}. In contrast, a fully electrodynamic picture for planar interfaces~\cite{moreau_prb56} or spherical nanoparticles (NPs) with aloof electrons followed much later~\cite{pogorzelski_pra8,Garcia1999}.
Electron trajectories penetrating the sphere are analytically and numerically more challenging, and only recently were such approaches finally developed for the case of homogeneous spheres~\cite{matsukata_nn15,Elli2024}.

Here, to further facilitate the study of light--matter coupling, which in most nanophotonic designs involves more than one material component ---one playing the role of the quantum emitter and the other that of the cavity~\cite{fofang_nl8,zengin_prl114,todisco_nanoph9,wang_jpcc124}, we extend the aforementioned solutions to the case of spherical core--shell NPs. One of the constituents can be a layer of organic J-aggregated molecules~\cite{bellessa_prl93} or a transition-metal dichalcogenide (TMD)~\cite{distefano_jpcc124}, thereby sustaining collective excitonic modes and enabling the realization of strong-coupling systems within a single NP, probed by a swift electron beam. We analyze the three different cases that one can encounter, i.e., aloof trajectory, electron traversing the shell only, or electron traveling through the full NP body, and derive the main expressions that need to be considered in each case. As examples of strong-coupling scenarios, we explore plasmonic--excitonic NPs~\cite{fofang_nl8}. Here, the ability to probe for strong coupling of the so-called plexcitons is found to only slightly depend on the primary parameters of the electron beam, i.e., electron velocity and impact parameter. We also study high-index dielectric--excitonic systems~\cite{Pres1}, where, in contrast, the strong coupling features of the Mie excitons can be selectively probed or masked by a corresponding choice of the position and energy of the incident electron. We thus provide a missing piece in the electron-beam spectroscopy toolbox, which we show has large potential for tailoring electron--optical--excitonic interactions.

\section{Theory}\label{sec:theory}

In this section, we present the analytical framework for describing the interaction of fast electron beams with spherical core--shell NPs. The theoretical approach is based on Mie theory, and builds on a framework recently reported for homogeneous spheres~\cite{Elli2024}.

\subsection{Aloof electron trajectory}

We first consider a spherical core–shell NP centered at the origin of the coordinate system. 
The particle consists of a core with radius $R_\mathrm{c}$, coated by a concentric shell of 
radius $R_\mathrm{s}$, and is surrounded by a host medium. The core, shell, and embedding 
medium are labeled as regions I, II, and III, respectively, and are characterized by 
relative permittivities $\varepsilon_i$ and relative permeabilities $\mu_i$, with $i=1,2,3$, 
respectively. Throughout all calculations presented here and in the following, the embedding 
medium is assumed to be air, with $\varepsilon_3=\mu_3=1$. Furthermore, all considered 
materials are non-magnetic, hence $\mu_i=1$ for all of them. As a consequence of the layered 
geometry, two material interfaces arise, namely an inner interface between the core and the 
shell, referred to as boundary A, and an outer interface between the shell and the 
surrounding medium, referred to as boundary B. 

The NP is excited by a fast electron beam, modeled as a single point charge $-e$ (where $e$ is the elementary charge), propagating with constant velocity $\mathbf{v} = v\mathbf{\hat{z}}$ along the $z$ axis. Its trajectory is given by $\mathbf{r}_\mathrm{e} = \mathbf{r}_0 + \mathbf{v}t$, with $\mathbf{r}_0 = (b,\phi_0,z_0 \to -\infty)$ in cylindrical coordinates, where the parameter $b$ denotes the impact parameter, defined as the distance between the electron trajectory and the axes origin, coinciding with the NP center.

\begin{figure*}[t]
\centering
\includegraphics[width=0.7\linewidth]{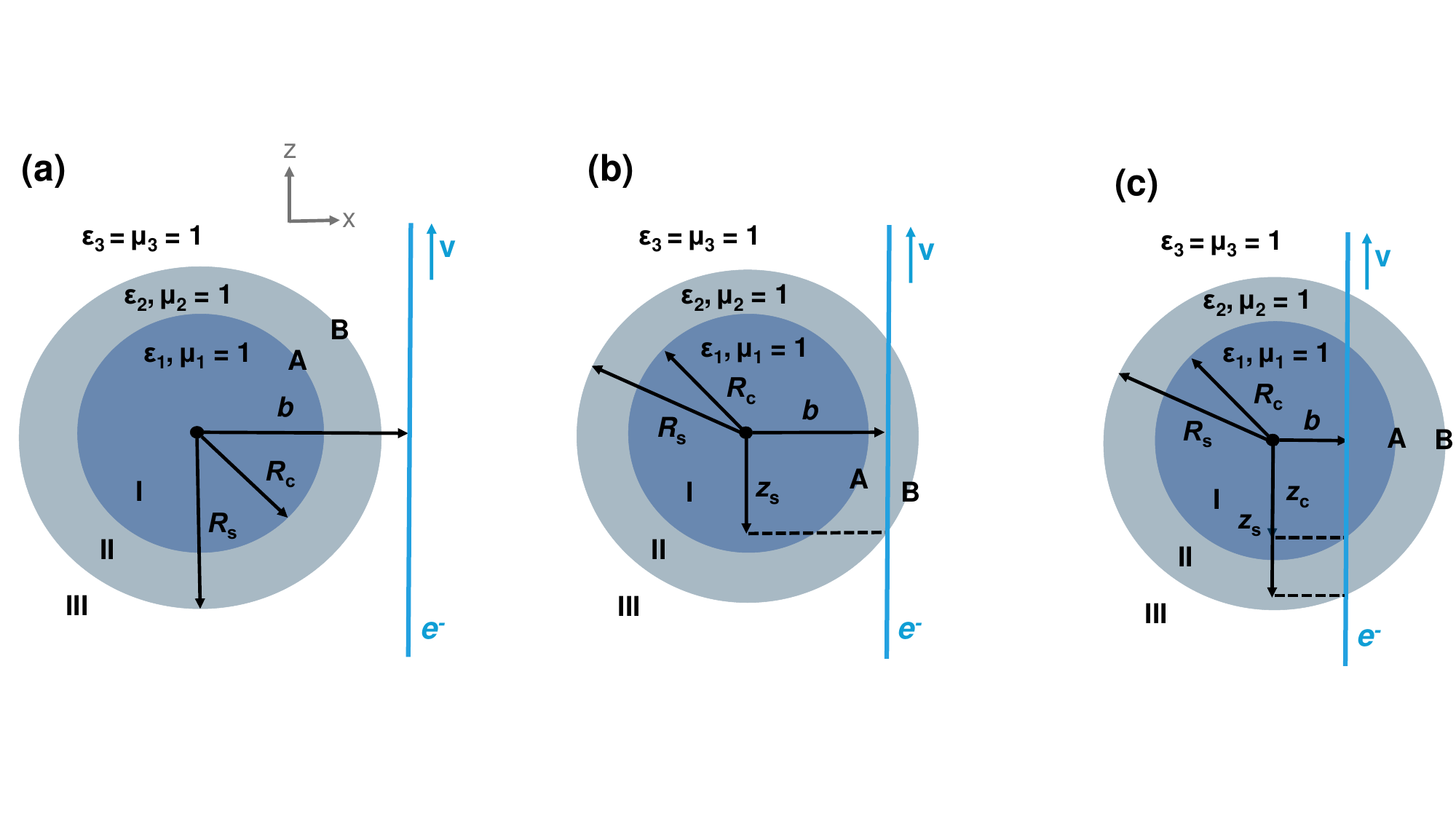}
\caption{Schematic illustration of an NP that consists of a core (region I) with radius $R_\text{c}$, surrounded by a shell (region II) of outer radius $R_\text{s}$, embedded in a host medium (region III). The three regions are characterized by their optical parameters $\varepsilon_i$ and $\mu_i=1$ with $i=1,2,3$. In all calculations, we consider air as the host medium with $\varepsilon_3=1$. An electron beam propagates parallel to the $z$–axis with impact parameter $b$. (a) Aloof trajectory: the electron remains entirely outside the NP ($b > R_\text{s}$). (b) Shell–penetrating trajectory: the electron enters the shell at $z=-z_\text{s}$ and exits at $z=z_\text{s}$ ($R_\text{s}>b>R_\text{c}$). (c) Core–penetrating trajectory: the electron additionally traverses the core, entering at $z=-z_\text{c}$ and exiting at $z=z_\text{c}$ ($R_\text{s}>R_\text{c}>b$).}
\label{fig:fig1}
\end{figure*}

We begin by considering electron beams passing by the nanosphere without intersecting it, i.e., on an aloof trajectory, as illustrated in Fig.~\ref{fig:fig1}(a). The resulting electromagnetic
(EM) fields in the frequency domain in each of the three regions can be expanded in the basis of spherical waves as~\cite{Stamatopoulou:25}
\begin{align}\label{eq:test}
& \mathbf{E}(\mathbf{r}) = \sum_{l,m}
\left[ b_{lm}^{\eta}h^+_l(k_ir)\mathbf{X}_{lm}(\mathbf{\hat{r}}) +
\frac{\mathrm{i}}{k_i}a_{lm}^{\eta}\nabla\times h^+_l(k_ir)\mathbf{X}_{lm}(\mathbf{\hat{r}} ) \right.\nonumber\\
& \left. + b_{lm}^{\zeta}j_l(k_ir)\mathbf{X}_{lm}(\mathbf{\hat{r}}) + 
\frac{\mathrm{i}}{k_i}a_{lm}^{\zeta}\nabla\times j_l(k_ir)\mathbf{X}_{lm}(\mathbf{\hat{r}} ) \right],
\end{align}
\begin{align}\label{eq:test2}
& \mathbf{H}(\mathbf{r})= \frac{1}{Z_i}\sum_{l,m} 
\bigg[ a_{lm}^{\eta}h^+_l(k_ir)\mathbf{X}_{lm}(\mathbf{\hat{r}}) 
\nonumber\\
&
- \frac{\mathrm{i}}{k_i}b_{lm}^{\eta}\nabla\times h^+_l(k_ir)\mathbf{X}_{lm}(\mathbf{\hat{r}} ) 
+ a_{lm}^{\zeta}j_l(k_ir)\mathbf{X}_{lm}(\mathbf{\hat{r}})
\nonumber\\
& - \frac{\mathrm{i}}{k_i}b_{lm}^{\zeta}\nabla\times j_l(k_ir)\mathbf{X}_{lm}(\mathbf{\hat{r}} ) \bigg],
\end{align}
where $i$=1, 2, and 3 denotes the three regions I, II, and III, and $k_i = \sqrt{\varepsilon_i\mu_i}\omega/c$ are the corresponding wavenumbers, where $c$ is the speed of light in vacuum. Here, $\zeta = $ I and $a_{l m}^{\eta} = b_{l m}^{\eta}=0$ in region I, $\eta = \mathrm{II,A}$ and $\zeta = \mathrm{II, B}$ in region II, while $\eta = \mathrm{III}$ and $\zeta = 0, \mathrm{III}$ in region III. Furthermore, $j_l(k_ir)$ denotes the spherical Bessel function, $h^+_l(k_ir)$ the spherical Hankel function of the first kind, $\mathbf{X}_{lm}(\mathbf{\hat{r}})$ are the vector spherical harmonics of degree $l$ and order $m$ depending on the angular direction $\mathbf{\hat{r}} = \theta, \phi$, and $Z_{i} = \sqrt{\mu_{i}\mu_0/(\varepsilon_{i}\varepsilon_0)}$ the wave impedance in medium $i$. We note that the notation $\sum_{l,m}$ denotes double summation over indices $l$ and $m$, with $l\leq l_\textrm{max}$ and $-l \leq m \leq l$.

The expansion coefficients $a_{lm}^\text{0,III}$ and $b_{lm}^\text{0,III}$ correspond to the incident field of the electron in region $\mathrm{III}$ and are given by Eqs.~\eqref{eq:alm0} and \eqref{eq:blm0} in Appendix~\ref{App:efield}. To obtain the coefficients $a_{lm}^\text{I}$ and $b_{lm}^\text{I}$ for the field inside the core, $a_{lm}^\text{II,A}$, $b_{lm}^\text{II,A}$, $a_{lm}^\text{II,B}$, and $b_{lm}^\text{II,B}$, for the field inside the shell associated with scattering at interfaces A and B, respectively, as well as $a_{lm}^\text{III}$ and $b_{lm}^\text{III}$ for the scattered field outside the NP, the EM interface conditions at interfaces A and B must be applied. Under the assumption that neither the core nor the shell sustains a surface current density, the tangential components of the electric and magnetic fields are continuous across both interfaces. We therefore find 
\begin{align}
\label{eq:ab_lm_I_aloof}
a/b_{lm}^\text{I} = T_{\text{E/M}l}^{21}a/b_{lm}^\text{II,B} = T_{\text{E/M}l}^{21}T_{\text{E/M}l}^{32}a/b_{lm}^\text{0,III} ,
\end{align}
\begin{align}
\label{eq:ab_lm_IIA_aloof}
a/b_{lm}^\text{II,A} = T_{\text{E/M}l}^{22}a/b_{lm}^\text{II,B} = T_{\text{E/M}l}^{22}T_{\text{E/M}l}^{32}a/b_{lm}^\text{0,III},
\end{align}
\begin{align}
\label{eq:ab_lm_IIB_aloof}
a/b_{lm}^\text{II,B} = T_{\text{E/M}l}^{32}a/b_{lm}^\text{0,III},
\end{align}
\begin{align}
\label{eq:ab_lm_III_aloof}
a/b_{lm}^\text{III} = T_{\text{E/M}l}^{33}a/b_{lm}^\text{0,III} ,
\end{align}
where $T_{\text{E/M}l}^{ij}$ with $i,j=1,2,3$ denote the Mie coefficients, which establish explicit relations between the multipole expansion coefficients of the incident, internal, and scattered fields. The exact expressions of the Mie coefficients are given in Appendix~\ref{App:Mie}.

With all field expansion coefficients fully determined, the EM response of the nanosphere to the passing electron is now completely described. The EELS and CL measurements can be simulated with the respective EEL and CL probabilities.
The EEL probability expresses the probability of the electron losing energy $\hbar\omega$ due to the interaction with the NP. It is related to the force acting on the electron by the induced field along its trajectory. In an aloof trajectory, the induced field corresponds precisely to the scattered field outside the NP. Using the scattered electric field in region $\textrm{III}$ of Eq.~\eqref{eq:test}, the total EEL probability for an aloof electron trajectory is obtained as~\cite{abajo_rmp82,Stamatopoulou:25}
\begin{align}
\label{eq:EEL_aloof_gesamt}    
& \Gamma_{\text{EEL}}(\omega) = -\frac{e^2}{\pi \varepsilon_0}\frac{1}{c\hbar\omega} \sum_{l,m}\frac{ \left| K_m \left( \frac{\omega b}{v\gamma_3} \right)\right| ^2}{l(l+1)} \times
\nonumber\\
& \left( \left|m\mathcal{M}^+_{lm}\right|^2\text{Re}\left[ T^{33}_{\text{M}l}\right] +  \left| \frac{\mathcal{N}^+_{lm}}{\beta\gamma_3} \right| ^2\text{Re}\left[ T^{33}_{\text{E}l}\right] \right).
\end{align}
Here, $K_m(x)$ is the modified Bessel function of the second kind, $T_{E/Ml}^{33}$ are Mie coefficients provided in Appendix~\ref{App:Mie}, while the $\mathcal{M}_{\ell m}^{+}$, $\mathcal{N}_{\ell m}^{+}$ coefficients can be found in Appendix~\ref{App:efield}.

The CL measurement is simulated using the CL probability, expressing the probability of detecting a photon of energy $\hbar\omega$ in the far field. It can be computed by integrating the Poynting flux of the induced EM field in the far field (at $r \to \infty$), in the direction normal to the surface.
Inserting the scattered EM fields in region $\textrm{III}$ of Eqs.~\eqref{eq:test} and \eqref{eq:test2} with the corresponding coefficients then yields
\begin{align}
\label{eq:CL_aloof_gesamt}    
& \Gamma_{\text{CL}}(\omega) = \frac{e^2}{\pi \varepsilon_0}\frac{1}{c\hbar\omega} \sum_{l,m}\frac{ \left| K_m \left( \frac{\omega b}{v\gamma_3} \right)\right| ^2}{l(l+1)} \times
\nonumber\\
& \left( \left|m\mathcal{M}^+_{lm}\right|^2 \left| T^{33}_{\text{M}l}\right|^2 +  \left| \frac{\mathcal{N}^+_{lm}}{\beta\gamma_3} \right| ^2 \left| T^{33}_{\text{E}l} \right|^2 \right).
\end{align}

\subsection{Core--shell penetrating electron trajectory}

We extend the analysis to penetrating trajectories, in which the electron beam traverses the NP and acts as a source both inside and outside the NP. We begin with the core–-shell penetrating configuration shown in Fig.~\ref{fig:fig1}(c), as it represents the most general case, from which the shell–penetrating geometry in Fig.~\ref{fig:fig1}(b) can be obtained as a special case. In the core-–shell penetrating configuration, the electron is assumed to enter the shell at position $(b,\phi_0,-z_\text{s} = -\sqrt{R_\text{s}^2-b^2})$ and the core at the position $(b,\phi_0,-z_\text{c} = -\sqrt{R_\text{c}^2-b^2})$, and to exit at $(b,\phi_0,z_\text{c})$ and $(b,\phi_0,z_\text{s})$, as shown in Fig.~\ref{fig:fig1}(c). 

As a consequence of the modified electron trajectory, the form of the EM fields in the different regions must be adjusted. At interface A, the fields are given by Eqs.~\eqref{eq:test} and \eqref{eq:test2}, with coefficients $a/b_{l m}^{\eta} = a/b_{l m}^{0,\mathrm{I}}$ and $a/b_{l m}^{\zeta} = a/b_{l m}^{\mathrm{I}}$ in region $\mathrm{I}$, $a/b_{l m}^{\eta} = a/b_{l m}^{\mathrm{II,A}} $ and $a/b_{l m}^{\zeta} = a/b_{l m}^{0,\mathrm{II,A}} +  a/b_{l m}^{\mathrm{II,B}}$ in region $\mathrm{II}$. At interface B, the expansion coefficients in region $\mathrm{II}$ take the form  $a/b_{l m}^{\eta} = a/b_{l m}^{\mathrm{II,A}} + a/b_{l m}^{0,\mathrm{II,B}}$ and $a/b_{l m}^{\zeta} =   a/b_{l m}^{\mathrm{II,B}}$, and $a/b_{l m}^{\eta} = a/b_{l m}^{\mathrm{III}}$ and $a/b_{l m}^{\zeta} = a/b_{l m}^{0,\mathrm{III}}$ in region $\mathrm{III}$.
The different representation of the coefficients in region $\mathrm{II}$ emerges because the expansion of the electron field takes the form of regular spherical waves at boundary A and singular spherical waves at boundary B inside the shell. 

To find the expansion coefficients $a/b_{lm}^\text{0,I}$, $a/b_{lm}^\text{0,II,A}$, $a/b_{lm}^\text{0,II,B}$, and $a/b_{lm}^\text{0,III}$ one has to account for the finite path length of the electron beam trajectory in the different regions as well as for the relation between $\mathbf{r}$ and the position of the electron $\mathbf{r}_e$. The explicit expressions for the respective coefficients can be found in Appendix \ref{App:efield2}. The remaining coefficients $a_{lm}^{\text{I/II(A,B)/III}}$ and $b_{lm}^{\text{I/II(A,B)/III}}$ of the induced fields are obtained by applying the interface conditions again at both interfaces $r=R_\text{c}$ and $r=R_\text{s}$ of the NP. The resulting expressions are given in Appendix \ref{appendix:exp_coeffs}.
The total EEL probability can then be written as
\begin{align}
\Gamma_\text{EEL}(\omega) &= 
\Gamma_\text{bulk}^\text{I}(\omega) + \Gamma_\text{bulk}^\text{II}(\omega) + 
\nonumber\\
& \Gamma_\text{begr}^\text{I}(\omega) + \Gamma_\text{begr}^\text{II,A}(\omega) + \Gamma_\text{begr}^\text{II,B}(\omega) + \Gamma_\text{surf}(\omega).
\end{align}

Here, the integral over the work done by the induced field is split into different contributions, whose explicit expressions can be found in Appendix~\ref{App:EEL}. The bulk ($\Gamma_\text{bulk}^\text{I}$, $\Gamma_\text{bulk}^\text{II}$) and Begrenzung ($\Gamma_\text{begr}^\text{I}$, $\Gamma_\text{begr}^\text{II,A}$, $\Gamma_\text{begr}^\text{II,B}$) terms contain the contributions of the electron beam trajectory that lie inside the NP. The bulk terms contain the contribution to the total EEL probability that stems from the propagation of the electron within the unbound media I and II, while the Begrenzung terms account for the presence of the interfaces A and B. The surface term $\Gamma_\text{surf}$ contains the contributions that arise from the excitation of the NP by the external part of the electron trajectory.
It is important to note that, in the absence of the NP, the expansion coefficients $a_{lm}^\text{I/II(A,B)/III}$, $b_{lm}^\text{I/II(A,B)/III}$ remain nonzero and therefore $\Gamma_\text{surf}(\omega)$ and $\Gamma_\text{begr}^\text{I/II(A,B)/III}(\omega)$ still contain contributions from the electron's direct field. As a result, $\Gamma_\text{EEL}(\omega)$ is nonzero even when no structure is present. To correct for this, the expansion coefficients are adjusted to \cite{Stamatopoulou:25}
\begin{align}
\label{eq:almIIA_corr}
a_{lm}^\text{II,A} = a_{lm}^\text{II,A} - a_{lm}^\text{0,I}\big|_{\varepsilon_1=\varepsilon_2},
\end{align}
\begin{align}
\label{eq:blmIIA_corr}
b_{lm}^\text{II,A} = b_{lm}^\text{II,A} - b_{lm}^\text{0,I}\big|_{\varepsilon_1=\varepsilon_2},
\end{align}
and 
\begin{align}
\label{eq:almIII_corr}
a_{lm}^\text{III} = a_{lm}^\text{III} - a_{lm}^\text{0,I}\big|_{\varepsilon_1=1} -a_{lm}^\text{0,II,B}\big|_{\varepsilon_2=1},
\end{align}
\begin{align}
\label{eq:blmIII_corr}
b_{lm}^\text{III} = b_{lm}^\text{III} - b_{lm}^\text{0,I}\big|_{\varepsilon_1=1} -b_{lm}^\text{0,II,B}\big|_{\varepsilon_2=1}.
\end{align}
The corresponding correction in the Begrenzung terms is already incorporated in the expressions given in Appendix~\ref{App:EEL}.

Finally, the CL probability can be obtained as
\begin{align}
\label{eq:GCL_pen_core}
\Gamma_{\text{CL}}(\omega) = \frac{1}{Z_3\pi\hbar\omega k_3^2}\sum_{l,m} \left( \left| a_{lm}^\text{III} \right|^2 + \left| b_{lm}^\text{III} \right|^2 \right) ,    
\end{align}
using the corrected expansion coefficients.

\subsection{Shell-penetrating electron trajectory}

We finally consider the case of an electron trajectory that penetrates the NP shell but does not enter the core, cf. Fig.~\ref{fig:fig1}(b). Compared to the fully penetrating core--shell geometry discussed in the previous subsection, this configuration represents a reduced case in which the electron acts as a source in the surrounding medium and within the shell only. This reduced case leads to modified expansion coefficients $a/b_{lm}^\text{0,II,A}$ and $a/b_{lm}^\text{0,II,B}$, which are obtained from the expressions derived for the fully penetrating case by replacing the integration bounds with the corresponding shell trajectory. Since the electron trajectory outside the NP is unchanged compared to the fully penetrating case, the incident–field coefficients in the surrounding medium, $a_{lm}^\text{0,III}$ and $b_{lm}^\text{0,III}$, remain identical. In contrast, no incident field is generated inside the core for the shell--penetrating trajectory, such that the corresponding expansion coefficients $a/b_{lm}^{\eta}$ vanish in Eqs.~\eqref{eq:test} and \eqref{eq:test2} for region $\textrm{I}$. 

Consequently, the expansion coefficients of the induced fields coincide with those listed in Appendix~\ref{appendix:exp_coeffs}, evaluated with $a_{lm}^{\text{0,I}} = b_{lm}^{\text{0,I}} = 0$. As in the core--shell penetrating case, the coefficients $a_{lm}^\text{III}$ and $b_{lm}^\text{III}$ are subsequently corrected according to Eqs.~\eqref{eq:almIII_corr} and \eqref{eq:blmIII_corr} to remove the contribution of the direct electron field and retain only the structure–induced response.

The EEL probability can be evaluated in full analogy to the penetrating core--shell case. The individual contributions, however, must be adapted to the modified electron trajectory within the shell region. In the present geometry, the total loss probability is given by
\begin{align}\label{eq:shell_decomp}
\Gamma_\text{EEL}(\omega) = \Gamma^\text{II}_\text{bulk}(\omega) + \Gamma^\text{II,A}_\text{begr}(\omega) + \Gamma^\text{II,B}_\text{begr}(\omega) + \Gamma_\text{surf}(\omega) .
\end{align}
The individual terms, as well as the modified expansion coefficients $a/b_{lm}^\text{0,II,A}$ and $a/b_{lm}^\text{0,II,B}$ can be found in Appendix \ref{app:EEL_shell_pen}.
The CL probability can still be obtained from Eq.~\eqref{eq:GCL_pen_core} using the corrected expansion coefficients $a_{lm}^\text{III}$ and $b_{lm}^\text{III}$.

\section{Results and discussion}

We now apply the method presented in Section~\ref{sec:theory} to probe strong coupling in CL and EEL spectra, and explore the impact of the electron beam parameters on the visibility of the strong coupling features. We focus on two types of systems: plasmon--exciton coupling in an exciton core--silver shell NP, and Mie--exciton coupling in a silicon core--exciton shell NP. We discuss them in two separate subsections.

\subsection{Plexcitons}

\begin{figure*}[t]
\centering
\includegraphics[width=\linewidth]{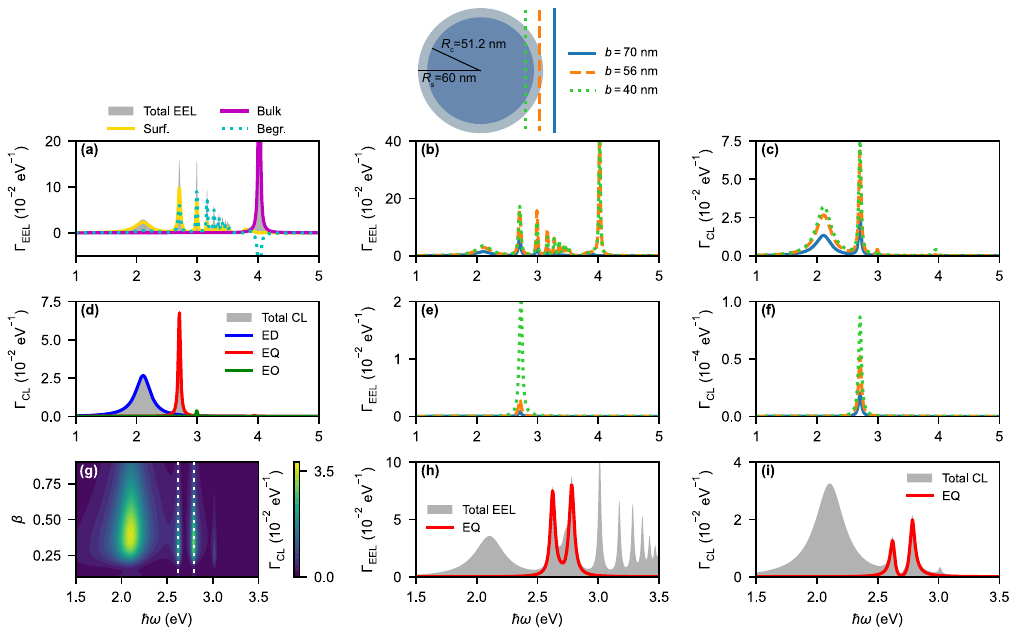}
\caption{(Top) Schematic of a silver core--exciton shell NP with $R_\mathrm{c}=51.2$\;nm and $R_\mathrm{s} = 60$\;nm. All line plots have been calculated with $v = 0.5c$. (a) EEL probability of a silver shell decomposed to surface, bulk, and Begrenzung contributions for $b = 40$\;nm. (b) EEL and (c) CL probability of the isolated silver shell for the different electron trajectories displayed in the schematic. (d) CL probability of a silver shell decomposed to its multipolar content---electric dipole (ED), electric quadrupole (EQ), and electric octupole (EO)---for $b = 40$\;nm. (e) EEL and (f) CL probability of the exciton core for different electron trajectories. (g) CL probability versus reduced velocity $\beta=v/c$ considering the core--shell NP for $b = 40$\;nm, The white dotted lines indicate the hybrid plexciton modes. (h) EEL and (i) CL probability along with the EQ contribution for the core--shell NP for $b=40$\;nm.}
\label{fig:fig2}
\end{figure*}

In this subsection, we investigate the plasmon–exciton coupling in spherical NPs consisting of an excitonic core surrounded by a silver shell, and excited by fast electron beams with varying impact parameter $b$ and velocity $v = 0.5c$, as illustrated in the top schematic of Fig.~\ref{fig:fig2}. The silver shell is modeled with a Drude dielectric function $\varepsilon(\omega) = \varepsilon_{\infty, \text{Ag}} - \omega_\textrm{p,Ag}^2/(\omega^2+ \mathrm{i} \omega \gamma_\text{Ag})$, with background permittivity $\varepsilon_{\infty,\text{Ag}}=5$, plasma energy $\hbar\omega_\text{p,Ag} = 8.99$\;eV, and damping $\hbar\gamma_\text{Ag} = 0.025$\;eV. The excitonic core is described with a Lorentz permittivity $\varepsilon (\omega) = \varepsilon_{\infty, \text{x}} - f\omega_\text{x}^2/(\omega^2-\omega_\text{x}^2 + \mathrm{i} \omega \gamma_\text{x})$, with background permittivity $\varepsilon_{\infty, \text{x}}=1$, oscillator strength $f=0.02$, transition energy $\hbar\omega_\text{x} = 2.7$\;eV, and damping $\hbar\gamma_{\text{x}}=0.052$\;eV. In all EEL calculations, we have used $q_\mathrm{c} = 30$\;nm$^{-1}$.

Panels~\ref{fig:fig2}(a) and \ref{fig:fig2}(d) show the EEL and CL probability of the isolated silver shell, for $b=56$\;nm, decomposed according to Eqs.~\eqref{eq:shell_decomp} and \eqref{eq:GCL_pen_core}. In the EEL spectrum, we observe a pronounced peak at energy $\hbar\omega_\text{p,Ag}/\sqrt{\varepsilon_{\infty,Ag}}=4.02$\;eV, corresponding to the excitation of a bulk plasmon. Furthermore, we observe the emergence of several peaks between $2.11$\;eV and $3.53$\;eV, corresponding to the excitation of localized surface plasmons (LSPs) of increasing orders. These modes are of electric type and accumulate at the equivalent surface plasmon polariton (SPP) frequency $\hbar\omega_\text{p,Ag}/\sqrt{1+\varepsilon_{\infty,Ag}}=3.67$\;eV for a planar silver--air interface. In the present calculation, the accumulation point appears at slightly lower energy due to the truncation of the multipole expansion at $l_\mathrm{max}=10$. At the same time, the silver shell sustains LSPs in both the inner (cavity-like LSPs) and outer metal--air interface (particle-like LSPs), leading to hybridization between the two and a small energy shift of the particle-like plasmon to lower energies and the cavity-like to higher energies~\cite{tserkezis_jpcm20}. This energy splitting becomes more pronounced as the shell thickness decreases. The cavity-like plasmons emerge close to the bulk plasmon energy, and are therefore obscured by the bulk plasmon peak. Contribution from modes with $l_\mathrm{max}>10$ is not negligible, but not relevant in the scope of this analysis. Only the three lowest orders of these modes are radiative, as revealed by the CL spectrum in Fig.~\ref{fig:fig2}(d). 

The total EELS and CL spectra for both isolated constituents, namely the silver shell and the excitonic shell, for different electron beam trajectories are presented in Figs.~\ref{fig:fig2}(b),\ref{fig:fig2}(c) and \ref{fig:fig2}(e),\ref{fig:fig2}(f). Here, the complementary region is considered to be air. The geometry of the silver shell is chosen such that its electric quadrupolar (EQ) mode is tuned at the exciton transition energy.
For the excitonic core, and for decreasing impact parameter, the peak intensity increases in both observables in panels \ref{fig:fig2}(e) and \ref{fig:fig2}(f), reflecting the stronger interaction between the electron beam and the core, while no other quantitative differences appear between different impact parameters. In contrast, the EEL spectrum of the silver shell in panel \ref{fig:fig2}(b) shows distinct behavior when comparing the aloof trajectory to the penetrating trajectories, since the bulk mode is excited only in the latter case. In the CL spectrum for $b=40$\;nm in panel \ref{fig:fig2}(c), we also observe a small peak at $3.96$\;eV, which corresponds to the cavity-like plasmons. Although this contribution is present in all CL and EEL spectra across different impact parameters, it is not always clearly distinguishable.

As our next step, we combine the two constituents in a core--shell NP. Theoretical studies of strong coupling in similar structures have been reported before, considering, however, only aloof trajectories~\cite{Zouros2020}. In Figs.~\ref{fig:fig2}(g) and \ref{fig:fig2}(h), we show the total EEL and CL spectra for a core--shell penetrating trajectory with $b=40$\;nm along with the contribution of the EQ mode, both revealing the splitting of the EQ mode in two peaks located at approximately $2.62$\;eV and $2.78$\;eV. Fitting the contribution of the EQ mode with a coupled harmonic oscillator model corroborates a coupling strength of $\hbar g = 0.16$\;eV, and therefore the strong coupling criterion $\hbar g > \sqrt{[(\hbar\gamma_\text{Ag})^2 + (\hbar\gamma_\text{x})^2]/2}$ is satisfied~\cite{Torma_2015}. Overall, the strong coupling signature is evident in both CL and EEL spectra, consistent with observations for aloof trajectories, revealing that the impact parameter is not an important factor in our ability to probe strong coupling.  

So far, we have analyzed the influence of different impact parameters and, thus, electron trajectories on the EEL and CL spectra. We now examine the effect of the electron beam velocity while maintaining the impact parameter fixed at $b=40$\;nm. Fig.~\ref{fig:fig2}(g) shows the total CL probability for the coupled core--shell NPs as a function of energy and reduced velocity $\beta=v/c$, with the white dotted lines marking the hybrid plexciton modes. We observe that the CL intensity increases with velocity, reaching a maximum for around $v=0.33c$ on average before decreasing at higher velocities. This trend is not specific to the coupled modes but applies equally to uncoupled resonances, while the characteristic features of the coupling remain unchanged with electron velocity.

Overall, the electron beam parameters do not play an important role in resolving the strong-coupling signature, neither in EEL nor in CL spectra, in the exciton core--silver shell configuration. These results may, at first glance, appear unsurprising: the coupling occurs between two characteristic eigenmodes of the system, which are, by definition, independent of the excitation. That is, unless the chosen source does not efficiently excite one of the modes, in which case the spectral signatures of the coupling may be strongly reduced or even absent. This scenario will be relevant in the next section, where we explore Mie excitons.  

\subsection{Mie excitons}

\begin{figure*}[t]
\centering
\includegraphics[width=\linewidth]{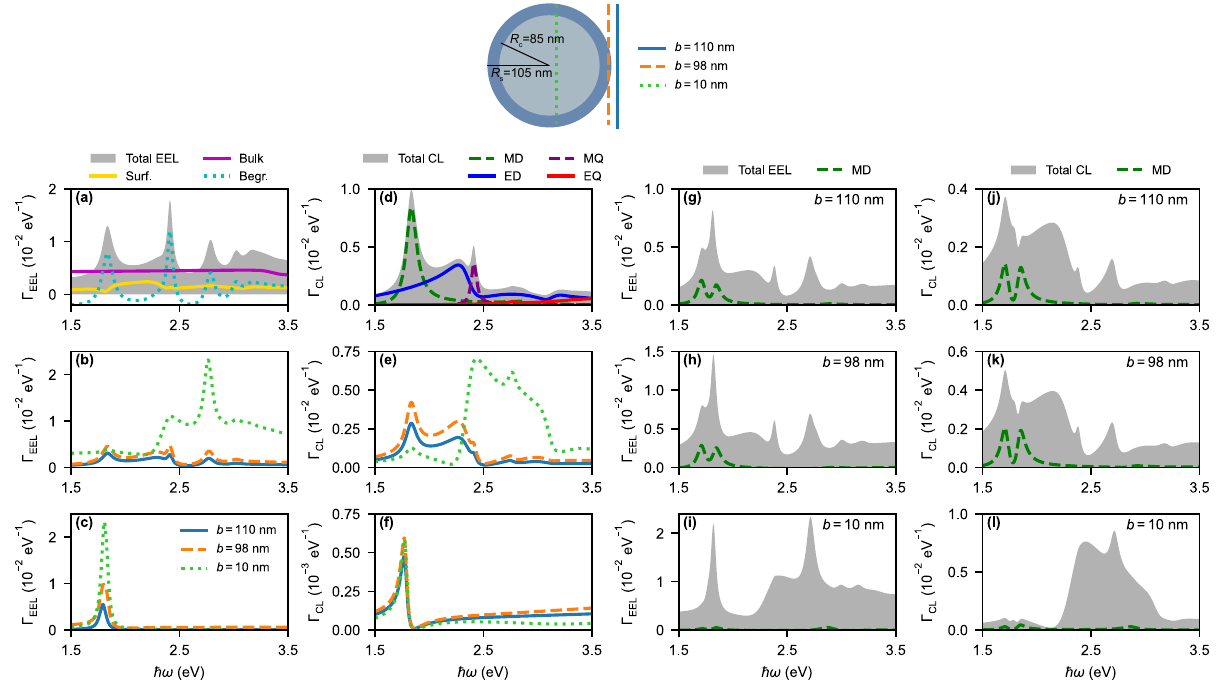}
\caption{(Top) Schematic of a silicon core--exciton shell NP with $R_\mathrm{c}=85$\;nm and $R_\mathrm{s} = 105$\;nm. All plots have been calculated with $v = 0.7c$. (a) EEL probability of a silicon NP decomposed to surface, bulk, and Begrenzung contributions for $b=10$\;nm. (b,c) EEL probability of (b) the isolated silicon core and (c) the exciton shell for the different electron trajectories displayed in the schematic. (d) CL probability of a silver shell decomposed to its multipolar content---magnetic dipole (MD), dashed green line; ED, solid blue line; EQ, solid red line; magnetic quadrupole (MQ), dashed purple line---for $b=10$\;nm. (e,f) CL probability of (e) the isolated silicon core and (d) the exciton shell for the different electron trajectories. (g-i) EEL and (j-l) CL probability along with the MD contribution for the core-shell NP for the different electron trajectories indicated in the inset.}
\label{fig:fig3}
\end{figure*}

We now turn to the coupling between Mie resonances supported by a silicon core of radius $R_\text{c}=85$\;nm and excitons hosted in a concentric shell of $R_\text{s}=105$\;nm, as illustrated in the schematic of Fig.~\ref{fig:fig3}. The silicon core is modeled with a frequency–dependent permittivity $\varepsilon_\text{Si}(\omega)$ given by experimental values~\cite{AspnesSi}. The excitonic shell is described by a Lorentz model with $\varepsilon_{\infty,\textrm{x}}=3$, $f=0.2$,  $\hbar\omega_\textrm{x} = 1.77$\;eV, and $\hbar \gamma_\textrm{x}=0.05$\;eV. For this calculation, the electron velocity is fixed to $v=0.7c$, and all calculations are performed using $q_\text{c}=2$\;nm$^{-1}$ and $l_\textrm{max}=5$.

Figs.~\ref{fig:fig3}(a) and \ref{fig:fig3}(d) show the respective EEL and CL spectra decomposed into the different contributions when probing the silicon core for $b=50$\;nm, revealing that Mie resonances of electric and magnetic type are sustained in the volume of the NP. The total EEL and CL probabilities considering a hollow excitonic shell for different electron trajectories are shown in Figs.~\ref{fig:fig3}(c) and \ref{fig:fig3}(f), exhibiting a resonance at the energy of the magnetic dipole (MD) mode of the silicon core. Whereas for the excitonic shell the impact parameter does not seem to affect significantly the excitation of the resonance, this is not the case for the silicon core in Figs.~\ref{fig:fig3}(b) and \ref{fig:fig3}(e), where significant differences appear in both observables when comparing the spectra of the penetrating trajectory to the aloof cases ($b=110$\;nm and $b=98$\;nm ). Such different spectral signatures arise because near-axis electron beams excite magnetic-type modes inefficiently. Further differences in the CL spectra between aloof and penetrating trajectories stem from the emergence of additional radiation channels when the electron travels through the NP, such as transition and Cherenkov radiation, with the latter being the dominant contribution within this energy and velocity range~\cite{Stamatopoulou:25,Fiedler2022}.

\begin{figure*}[t]
\centering
\includegraphics[width=\linewidth]{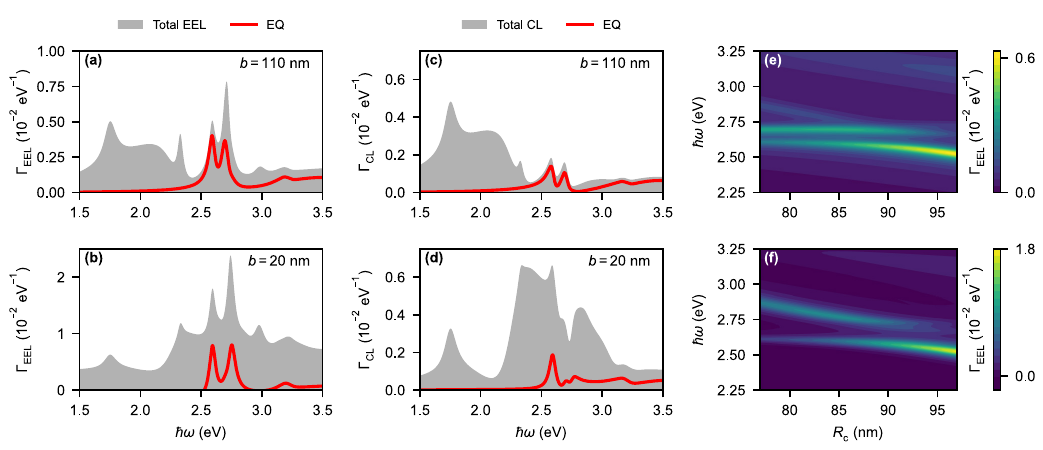}
\caption{Strong coupling in a silicon core--exciton shell NP with $R_\mathrm{c}=87$\;nm and $R_\mathrm{s} = 105$\;nm. All plots have been calculated with $v = 0.7c$. (a,b) EEL  probability along with the EQ contribution for (a) an aloof and (b) a penetrating electron trajectory. (c,d) CL probability along with the EQ contribution for (c) an aloof and (d) a penetrating electron trajectory. (e,f) EEL probability versus core radius $R_\mathrm{c}$ showing the anticrossing between the EQ mode and the exciton-polaritons for (e) an aloof and (f) a penetrating electron trajectory.}
\label{fig:fig4}
\end{figure*}

Having analyzed the isolated cores and shells, we now combine them into core--shell NPs to investigate the resolution of the strong coupling features, in Figs.~\ref{fig:fig3}(g)-\ref{fig:fig3}(l), showing the total EEL and CL probability along with the MD contribution for different electron trajectories. We note that the MD contribution includes only the surface and Begrenzung terms; the absence of the bulk contribution may result in negative values. For both the aloof ($b=110$\;nm) and the shell–penetrating trajectory ($b=98$\;nm), in Figs.~\ref{fig:fig3}(g)-\ref{fig:fig3}(h) and Figs.~\ref{fig:fig3}(j)-\ref{fig:fig3}(k), a clear double–peak structure is visible in both the EEL and CL spectra in the MD contribution. The two hybridized peaks of the coupled system are located at around $1.71$\;eV and $1.84$\;eV. Fitting the MD contribution with a coupled harmonic oscillator model yields coupling strengths of $\hbar g=0.13$\;eV, which satisfies the strong coupling criterion. Although the coupled mode does not dominate the overall spectra, the double–peak structure is still weakly reflected in the total probabilities for both trajectories. In contrast, when the electron beam penetrates the NP on a near–central trajectory in Figs.~\ref{fig:fig3}(i) and \ref{fig:fig3}(l), the double peak indicating the coupling is largely suppressed in both the EEL and CL spectra. Such a suppression is a direct consequence of the inefficient excitation of magnetic-type modes for near–axis electron trajectories.

On the other hand, electric-type Mie resonances can be efficiently excited for all impact parameters. In Fig.~\ref{fig:fig4}, we explore whether the strong coupling features between electric-type modes and electronic transitions can be resolved for all electron beam trajectories. To this end, we consider a modified system with core radius $R_\textrm{c} = 87$\,nm and shell $R_\textrm{s} = 105$\,nm and excitonic material with $\hbar\omega_\textrm{x} = 2.65$\;eV, while all other parameters remain unchanged. Figs.~\ref{fig:fig4}(a) and \ref{fig:fig4}(b) show the total EEL probability of the core--shell NP for an aloof ($b=110$\;nm) and a core--shell penetrating trajectory ($b=20$\;nm), along with the contribution of the EQ coupled mode. The double-peak feature is clearly visible not only in the isolated EQ contribution but also in the total EEL spectra, for both trajectories. In contrast, the CL spectrum for the penetrating trajectory in Fig.~\ref{fig:fig4}(d) does not provide an unambiguous signature of strong coupling. A substantial contribution from Cherenkov radiation slightly blueshifts the position of the peak corresponding to the EQ mode, obscuring the characteristic double peak. 

By comparing the EQ contribution in the EEL spectra in Figs.~\ref{fig:fig4}(a) and \ref{fig:fig4}(b), we observe that the splitting between the peaks associated with the hybrid modes is larger for the penetrating electron trajectory compared to the aloof case. Such an increased splitting does not indicate a larger coupling strength, but instead exposes the coupling of a different exciton-polariton branch with the EQ mode. This behavior can be better understood by analyzing the anticrossing branches of the EQ mode and exciton coupling in Figs.~\ref{fig:fig4}(e) and \ref{fig:fig4}(f). In both panels, three distinct branches emerge while varying the core radius, indicating that at least two exciton polaritons participate in the coupling. It can be seen that coupling with the middle branch is favored with the aloof electron trajectory, while the upper branch is more efficiently excited with a penetrating beam. 

To conclude, we find that in the case of Mie--exciton coupling, the electron beam parameters play a crucial role in resolving the strong-coupling signature. This sensitivity arises from the selective excitation of optical modes by the electron beam, as well as from the emergence of additional loss mechanisms when the beam intersects the structure.

\section{Conclusions}

We examined the interaction of fast electrons with strongly-coupled core--shell NPs, focusing on how the electron beam parameters, namely the impact parameter and velocity, affect our ability to probe the coupling. To this end, we devised an analytical framework for calculating EEL and CL probabilities in spherical core--shell NPs excited by fast electrons, based on Mie theory. The framework was applied to systems consisting of a plasmonic or dielectric nanocavity and an excitonic component. In plasmonic configurations, strong coupling between the optical mode and excitonic transitions leads to clear spectral mode splitting that remains robust against variations in beam parameters. In contrast, dielectric configurations exhibit a pronounced dependence on excitation conditions. On the one hand, magnetic Mie modes are weakly excited for near-axis electron trajectories, suppressing the spectral features associated with strong coupling. On the other hand, electric modes remain accessible for all impact parameters, but can be spectrally distorted in CL due to additional radiation channels. EEL spectra consistently show the signature features of strong coupling, albeit the electron beam position determines which exciton-polariton branch is preferentially excited. Our work provides a missing analytical formalism that can further clarify the origins and mechanisms of coupling in complex NPs, thereby accelerating the quest for miniaturized photonic architectures for quantum technologies.

\begin{acknowledgments}
P.~E.~S. is supported by a research grant (VIL71383) from VILLUM FONDEN. The Center for Polariton-driven Light--Matter Interactions (POLIMA) is sponsored by the Danish National Research Foundation (Project No. DNRF165).
\end{acknowledgments}


\section{Appendix}

\subsection{Electron-field expansion coefficients in an infinite homogeneous medium}\label{App:efield}

The coefficients that enter into the expansion of the field of the electron propagating in an infinite homogeneous medium of relative permittivity $\varepsilon$ and permeability $\mu$ with initial position $\mathbf{r}_0 = (b, \phi_0, z_0 \to \infty)$ are given by~\cite{abajo_rmp82,Stamatopoulou:25}
\begin{align}
\label{eq:alm0}
a_{lm}^{0,\mathrm{III}} = \frac{\mathrm{i} k^2e}{\varepsilon\varepsilon_0\omega} \frac{1}{\sqrt{\varepsilon\mu}\beta\gamma} \frac{\mathcal{N}^+_{lm}}{\sqrt{l(l+1)}}K_m\left( \frac{\omega b}{v\gamma} \right) \text{e}^{-\mathrm{i} m\phi_0} ,
\end{align}
and
\begin{align}
\label{eq:blm0}
b_{lm}^{0,\mathrm{III}} = -\frac{\mathrm{i} k^2e}{\varepsilon\varepsilon_0\omega}\frac{m\mathcal{M}^+_{lm}}{\sqrt{l(l+1)}}K_m\left( \frac{\omega b}{v\gamma} \right) \text{e}^{-\mathrm{i} m\phi_0},
\end{align}
where $k =\sqrt{\varepsilon\mu}\omega/c$, $\gamma = 1/\sqrt{1-\varepsilon\mu\beta^2}$ with $\beta = v/c$. In Eq.~\eqref{eq:blm0}, we have set
\begin{align}
\label{eq:M_pm}
\mathcal{M}_{lm}^+ = \mathrm{i}^{l+m}\sqrt{\frac{2l+1}{\pi}\frac{(l-m)!}{(l+m)!}} \frac{(2m-1)!!}{(\sqrt{\epsilon\mu}\gamma\beta)^m} G^{m+\frac{1}{2}}_{l-m}\left(\frac{1}{\sqrt{\varepsilon\mu}\beta}\right),
\end{align}
which holds for $m \geq 0$, while $\mathcal{M}_{l-m}^+ = (-1)^m\mathcal{M}_{lm}^+$, and $G_{\ell - m}^{m+1/2}(x)$ is the Gegenbauer polynomial.
In Eq.~\eqref{eq:alm0} we have set
\begin{align}
\label{eq:alm}
& \mathcal{N}^\pm_{lm} = \alpha_l^m\mathcal{M}^+_{l\ m+1} - \alpha_l^{-m}\mathcal{M}^+_{l\ m-1} \nonumber\\
& \text{with} \quad \alpha_l^m = \frac{1}{2}\sqrt{(l-m)(l+m+1)}.
\end{align}

\begin{widetext}
\subsection{Mie coefficients}\label{App:Mie}
For a core--shell NP, the Mie coefficients $T_{\text{E}l}^{ij}$ and $T_{\text{M}l}^{ij}$, with $i,j=1,2,3$, which relate the incident, internal, and scattered multipole coefficients in the three regions, are given by \cite{BohrenHuffman1983}
\begin{align}
\label{eq:TEl22}
T_{\text{E}l}^{22} = \frac{a^\text{II,A}_{lm}}{a_{lm}^\text{II,B}} = \frac{\varepsilon_1j_l(k_1R_\text{c})\Psi_l'(k_2R_\text{c})-\varepsilon_2\Psi_l'(k_1R_\text{c})j_l(k_2R_\text{c})}{\varepsilon_2h^+_l(k_2R_\text{c})\Psi_l'(k_1R_\text{c})-\varepsilon_1\xi_l'(k_2R_\text{c})j_l(k_1R_\text{c})} ,
\end{align}
\begin{align}
\label{eq:TMl22}
T_{\text{M}l}^{22} = \frac{b^\text{II,A}_{lm}}{b_{lm}^\text{II,B}} = \frac{j_l(k_1R_\text{c})\Psi_l'(k_2R_\text{c})-\Psi_l'(k_1R_\text{c})j_l(k_2R_\text{c})}{h^+_l(k_2R_\text{c})\Psi_l'(k_1R_\text{c})-\xi_l'(k_2R_\text{c})j_l(k_1R_\text{c})} ,
\end{align}
\begin{align}
\label{eq:TEl21}
T_{\text{E}l}^{21} = \frac{a^\text{I}_{lm}}{a_{lm}^\text{II,B}} =
\frac{\varepsilon_2k_1}{k_2} \frac{h^+_l(k_2R_\text{c})\Psi_l'(k_2R_\text{c})-\xi_l'(k_2R_\text{c})j_l(k_2R_\text{c})}{\varepsilon_2h^+_l(k_2R_\text{c})\Psi_l'(k_1R_\text{c})-\varepsilon_1\xi_l'(k_2R_\text{c})j_l(k_1R_\text{c})} ,
\end{align}
\begin{align}
\label{eq:TMl21}
T_{\text{M}l}^{21} = \frac{b^\text{I}_{lm}}{b_{lm}^\text{II,B}} =
\frac{h^+_l(k_2R_\text{c})\Psi_l'(k_2R_\text{c})-\xi_l'(k_2R_\text{c})j_l(k_2R_\text{c})}{h^+_l(k_2R_\text{c})\Psi_l'(k_1R_\text{c})-\xi_l'(k_2R_\text{c})j_l(k_1R_\text{c})} ,
\end{align}
\begin{align}
\label{eq:TEl33}
T_{\text{E}l}^{33} = \frac{a^\text{III}_{lm}}{a_{lm}^\text{0,III}} =  
\frac{(\varepsilon_2h^+_l(x_2)\Psi_l'(x_3)-\xi_l'(x_2)j_l(x_3))T^{22}_{\text{E}l} + \varepsilon_2j_l(x_2)\Psi'_l(x_3)-\Psi'_l(x_2)j_l(x_3)}{(h^+_l(x_3)\xi_l'(x_2)-\varepsilon_2\xi_l'(x_3)h^+_l(x_2))T^{22}_{\text{E}l} + h^+_l(x_3)\Psi'_l(x_2)-\varepsilon_2\xi'_l(x_3)j_l(x_2)} ,
\end{align}
\begin{align}
\label{eq:TMl33}
T_{\text{M}l}^{33} =\frac{b^\text{III}_{lm}}{b_{lm}^\text{0,III}} =  \frac{(h^+_l(x_2)\Psi_l'(x_3)-\xi_l'(x_2)j_l(x_3))T^{22}_{\text{E}l} + j_l(x_2)\Psi'_l(x_3)-\Psi'_l(x_2)j_l(x_3)}{(h^+_l(x_3)\xi_l'(x_2)-\xi_l'(x_3)h^+_l(x_2))T^{22}_{\text{E}l} + h^+_l(x_3)\Psi'_l(x_2)-\xi'_l(x_3)j_l(x_2)} ,
\end{align}
\begin{align}
\label{eq:TEl32}
T_{\text{E}l}^{32} =\frac{a^\text{II,B}_{lm}}{a_{lm}^\text{0,III}} = \frac{1}{x_3} \frac{-i\sqrt{\varepsilon_2}}{(h^+_l(x_3)\xi_l'(x_2)-\varepsilon_2\xi_l'(x_3)h^+_l(x_2))T^{22}_{\text{E}l} + h^+_l(x_3)\Psi'_l(x_2)-\varepsilon_2\xi'_l(x_3)j_l(x_2)} \ ,
\end{align}
\begin{align}
\label{eq:TMl32}
T_{\text{M}l}^{32} =\frac{b^\text{II,B}_{lm}}{b_{lm}^\text{0,III}} = \frac{1}{x_3} \frac{-i}{(h^+_l(x_3)\xi_l'(x_2)-\xi_l'(x_3)h^+_l(x_2))T^{22}_{\text{E}l} + h^+_l(x_3)\Psi'_l(x_2)-\xi'_l(x_3)j_l(x_2)} \ ,
\end{align}
with $x_{2,3} = k_{2,3}R_\text{s}$, where $\Psi_l(x) = xj_l(x)$ and $\xi_l(x) = xh_l^+(x)$ denote the Riccati–Bessel functions, and the prime indicates the derivative with respect to their argument.

\subsection{Electron-field expansion coefficients for a core--shell penetrating electron beam}\label{App:efield2}

We consider a core--shell penetrating configuration, where the electron is assumed to enter the shell at position $(b,\phi_0,-z_\text{s} = -\sqrt{R_\text{s}^2-b^2})$ and the core at the position $(b,\phi_0,-z_\text{c} = -\sqrt{R_\text{c}^2-b^2})$, and to exit symmetrically at $(b,\phi_0,z_\text{c})$ and $(b,\phi_0,z_\text{s})$. The explicit expansion coefficients read
\begin{align}
\label{eq:alm0I_pen_core-shell}
a_{lm}^\text{0,I} = \frac{\mathrm{i} k_1^2e}{\varepsilon_1\varepsilon_0\omega}\frac{\text{e}^{-\mathrm{i} m\phi_0}}{\sqrt{l(l+1)}} \frac{\mathrm{i}}{b} \int_{-z_\text{c}}^{z_\text{c}}\text{d}z \ \text{e}^{\mathrm{i}\omega z/v} (\mathcal{J}^+_{lm}(k_1z) + \mathcal{J}^-_{lm}(k_1z)) ,
\end{align}
\begin{align}
\label{eq:blm0I_pen_core-shell}
b_{lm}^\text{0,I} = -\frac{\mathrm{i} k_1^2e}{\varepsilon_1\varepsilon_0\omega}\frac{m\text{e}^{-\mathrm{i} m\phi_0}}{\sqrt{l(l+1)}}  \mathrm{i} k_1 \int_{-z_\text{c}}^{z_\text{c}}\text{d}z \ \text{e}^{\mathrm{i}\omega z/v}j_l(k_1r)Y_l^m(\theta,0) ,
\end{align}
\begin{align}
\label{eq:alm0IIA_pen_core-shell}
a_{lm}^\text{0,II,A} = \frac{\mathrm{i} k_2^2e}{\varepsilon_2\varepsilon_0\omega} \frac{e^{-\mathrm{i} m\phi_0}}{\sqrt{l(l+1)}} \frac{\mathrm{i}}{b} &\bigg(\int_{-z_\text{s}}^{z_\text{s}} \text{d}z \ e^{\mathrm{i}\omega z/v}(\mathcal{H}_{lm}^+(k_2z)+\mathcal{H}_{lm}^-(k_2z)) 
- \int_{-z_\text{c}}^{z_\text{c}} \text{d}z \ e^{\mathrm{i}\omega z/v}(\mathcal{H}_{lm}^+(k_2z)+\mathcal{H}_{lm}^-(k_2z)) \bigg) ,
\end{align}
\begin{align}
\label{eq:blm0IIA_pen_core-shell}
b_{lm}^\text{0,II,A} =  -\frac{\mathrm{i} k_2^2e}{\varepsilon_2\varepsilon_0\omega} \frac{me^{-\mathrm{i} m\phi_0}}{\sqrt{l(l+1)}} \mathrm{i} k_2 &\bigg(\int_{-z_\text{s}}^{z_\text{s}} \text{d}z \ e^{\mathrm{i}\omega z/v}h^+_l(k_2z)Y_l^m(\theta,0)
- \int_{-z_\text{c}}^{z_\text{c}} \text{d}z \ e^{\mathrm{i}\omega z/v}h^+_l(k_2z)Y_l^m(\theta,0) \bigg) ,
\end{align}
\begin{align}
\label{eq:alm0IIB_pen_core-shell}
a_{lm}^\text{0,II,B} =  \frac{\mathrm{i} k_2^2e}{\varepsilon_2\varepsilon_0\omega} \frac{e^{-\mathrm{i} m\phi_0}}{\sqrt{l(l+1)}} \frac{\mathrm{i}}{b} &\bigg(\int_{-z_\text{s}}^{z_\text{s}} \text{d}z \ e^{\mathrm{i}\omega z/v}(\mathcal{J}_{lm}^+(k_2z)+\mathcal{J}_{lm}^-(k_2z)) 
- \int_{-z_\text{c}}^{z_\text{c}} \text{d}z e^{\mathrm{i}\omega z/v}(\mathcal{J}_{lm}^+(k_2z)+\mathcal{J}_{lm}^-(k_2z)) \bigg) ,
\end{align}
\begin{align}
\label{eq:blm0IIB_pen_core-shell}
b_{lm}^\text{0,II,B} =  -\frac{\mathrm{i} k_2^2e}{\varepsilon_2\varepsilon_0\omega} \frac{me^{-\mathrm{i} m\phi_0}}{\sqrt{l(l+1)}} \mathrm{i} k_2 &\bigg(\int_{-z_\text{s}}^{z_\text{s}} \text{d}z \ e^{\mathrm{i}\omega z/v}j_l(k_2z)Y_l^m(\theta,0)
- \int_{-z_\text{c}}^{z_\text{c}} \text{d}z \ e^{\mathrm{i} \omega z/v}j_l(k_2z)Y_l^m(\theta,0) \bigg) ,
\end{align}
\begin{align}
\label{eq:alm0III_pen_core-shell}
a_{lm}^\text{0,III} = \frac{\mathrm{i} k_3^2e}{\varepsilon_0\omega}\frac{\text{e}^{-\mathrm{i} m\phi_0}}{\sqrt{l(l+1)}} \bigg( &\frac{\mathcal{N}^+_{lm}}{\beta\gamma_3} K_m\left( \frac{\omega b}{v\gamma_3} \right) 
- \frac{\mathrm{i}}{b} \int_{-z_\text{s}}^{z_\text{s}}\text{d}z \ \text{e}^{\mathrm{i}\omega z/v} (\mathcal{H}^+_{lm}(k_3z) + \mathcal{H}^-_{lm}(k_3z)) \bigg) ,
\end{align}
\begin{align}
\label{eq:blm0III_pen_core-shell}
b_{lm}^\text{0,III} = -\frac{\mathrm{i} k_3^2e}{\varepsilon_0\omega}\frac{m\text{e}^{-\mathrm{i} m\phi_0}}{\sqrt{l(l+1)}} \left( \mathcal{M}^+_{lm}K_m\left( \frac{\omega b}{v\gamma_3} \right) - \mathrm{i} k_3 \int_{-z_\text{s}}^{z_\text{s}}\text{d}z \ \text{e}^{\mathrm{i}\omega z/v}h^+_l(k_3r)Y_l^m(\theta,0)\right) ,
\end{align}
where $r=\sqrt{b^2+z^2}$, $\theta=\arccos{(z/r)}$, and $Y_l^m(\theta,0)$ are the scalar spherical harmonics. Here, we have set 
\begin{align}
\label{eq:Fpm}
& \mathcal{F}^\pm_{lm}(k_iz) =
\nonumber\\
& \mp \alpha_l^{\pm m} \Big[ (1\pm m) f_l(k_ir)Y_l^{m\pm 1}(\theta,0) + \frac{k_ib^2}{r} f_l'(k_ir)Y_l^{m\pm 1}(\theta,0)  \pm \frac{zb}{r^2}f_l(k_ir) (\alpha_l^{\pm m+1}Y_l^{m\pm 2}(\theta,0) - \alpha_l^{\pm m}Y_l^m(\theta,0)) \Big] ,
\end{align}
where $\mathcal{F}(x) = \mathcal{H}(x), \mathcal{J}(x)$ accounts for equations with $f_l(x) = h^+_l(x), j_l(x)$ accordingly, and the index $i=1,2,3$ denotes the respective material.

\subsection{Expansion coefficients of induced fields for core--shell NPs}
\label{appendix:exp_coeffs}

The explicit expressions of the expansion coefficients of the induced fields for a core--shell NP, obtained from the boundary conditions at the interfaces A and B, read

\begin{equation}
\label{eq:almI_pen_core-shell}
\begin{split}
a_{lm}^\text{I} = &
(Z_2(h^+_l(k_3R_\text{s})k_3 
(a_{lm}^\text{0,II,B}  k_1  (-h^+_l(k_2R_\text{c}) \Psi'_l(k_2R_\text{c}) + 
j_l(k_2R_\text{c}) \xi'_l(k_2R_\text{c})) \xi'_l(k_2R_\text{s}) 
+ 
a_{lm}^\text{0,I} k_2 \xi'_l(k_1R_\text{c}) (-h^+_l(k_2R_\text{c}) \Psi'_l(k_2R_\text{s}) \\
& + j_l(k_2R_\text{c}) \xi'_l(k_2R_\text{s}))) Z_1 + 
a_{lm}^\text{0,III} k_1 k_2 (h^+_l(k_2R_\text{c}) \Psi'_l(k_2R_\text{c}) - 
j_l(k_2R_\text{c}) \xi'_l(k_2R_\text{c})) (h^+_l(k_3R_\text{s}) \Psi'_l(k_3R_\text{s}) \\ 
&- j_l(k_3R_\text{s}) \xi'_l(k_3R_\text{s})) Z_1 + 
a_{lm}^\text{0,I} h^+_l(k_1R_\text{c}) h^+_l(k_3R_\text{s}) k_1 k_3 (\Psi'_l(k_2R_\text{s}) \xi'_l(k_2R_\text{c})
- \Psi'_l(k_2R_\text{c}) \xi'_l(k_2R_\text{s})) Z_2) + \\ 
& k_2 \xi'_l(k_3R_\text{s}) (a_{lm}^\text{0,I} (-h^+_l(k_2R_\text{s}) j_l(k_2R_\text{c})
+ h^+_l(k_2R_\text{c})j_l(k_2R_\text{s})) k_2 \xi'_l(k_1R_\text{c}) Z_1 + 
a_{lm}^\text{0,II,B} h^+_l(k_2R_\text{s}) k_1 (h^+_l(k_2R_\text{c}) \Psi'_l(k_2R_\text{c}) \\
&- j_l(k_2R_\text{c}) \xi'_l(k_2R_\text{c})) Z_1 + 
a_{lm}^\text{0,I} h^+_l(k_1R_\text{c}) k_1 (h^+_l(k_2R_\text{s}) \Psi'_l(k_2R_\text{c})
- j_l(k_2R_\text{s}) \xi'_l(k_2R_\text{c})) Z_2) Z_3 
\\
& + a_{lm}^\text{0,II,A} k_1 (h^+_l(k_2R_\text{c}) \Psi'_l(k_2R_\text{c})
- j_l(k_2R_\text{c}) \xi'_l(k_2R_\text{c})) Z_1 (h^+_l(k_3R_\text{s}) k_3 \Psi'_l(k_2R_\text{s}) Z_2 - 
j_l(k_2R_\text{s}) k_2 \xi'_l(k_3R_\text{s}) Z_3))\\
&/(h^+_l(k_3R_\text{s}) k_3 Z_2 (h^+_l(k_2R_\text{c}) k_2 \Psi'_l(k_1R_\text{c}) \Psi'_l(k_2R_\text{s}) Z_1 
-j_l(k_2R_\text{c}) k_2 \Psi'_l(k_1R_\text{c}) \xi'_l(k_2R_\text{s}) Z_1 + \\
& j_l(k_1R_\text{c}) k_1 (-\Psi'_l(k_2R_\text{s}) \xi'_l(k_2R_\text{c}) 
+ \Psi'_l(k_2R_\text{c}) \xi'_l(k_2R_\text{s})) Z_2) + 
k_2 \xi'_l(k_3R_\text{s}) (h^+_l(k_2R_\text{s}) j_l(k_2R_\text{c}) k_2 \Psi'_l(k_1R_\text{c}) Z_1 \\
&- h^+_l(k_2R_\text{c}) j_l(k_2R_\text{s}) k_2 \Psi'_l(k_1R_\text{c}) Z_1 - h^+_l(k_2R_\text{s}) j_l(k_1R_\text{c}) k_1 \Psi'_l(k_2R_\text{c}) Z_2
+ j_l(k_1R_\text{c}) j_l(k_2R_\text{s}) k_1 \xi'_l(k_2R_\text{c}) Z_2) Z_3),
\end{split}
\end{equation}
\begin{equation}
\label{eq:almIIA_pen_core-shell}
\begin{split}
a_{lm}^\text{II,A} = &
( Z_2  (a_{lm}^\text{0,III}  k_2  (h^+_l(k_3R_\text{s}) \Psi'_l(k_3R_\text{s}) - j_l(k_3R_\text{s}) \xi'_l(k_3R_\text{s}))  (j_l(k_2R_\text{c})   k_2 \Psi'_l(k_1R_\text{c})   Z_1
- j_l(k_1R_\text{c})   k_1 \Psi'_l(k_2R_\text{c})   Z_2) + \\
& h^+_l(k_3R_\text{s})   k_3  (a_{lm}^\text{0,I}  k_2 \Psi'_l(k_2R_\text{s})  (-h^+_l(k_1R_\text{c}) \Psi'_l(k_1R_\text{c})
+ j_l(k_1R_\text{c}) \xi'_l(k_1R_\text{c}))   Z_2 + 
a_{lm}^\text{0,II,B}\xi'_l(k_2R_\text{s})  (-j_l(k_2R_\text{c})   k_2 \Psi'_l(k_1R_\text{c})   Z_1\\
&+ j_l(k_1R_\text{c})   k_1 \Psi'_l(k_2R_\text{c})   Z_2))) + 
k_2 \xi'_l(k_3R_\text{s})  (a_{lm}^\text{0,I} j_l(k_2R_\text{s})   k_2  (h^+_l(k_1R_\text{c}) \Psi'_l(k_1R_\text{c})
- j_l(k_1R_\text{c}) \xi'_l(k_1R_\text{c}))   Z_2 + \\
& a_{lm}^\text{0,II,B} h^+_l(k_2R_\text{s})  (j_l(k_2R_\text{c})   k_2 \Psi'_l(k_1R_\text{c})   Z_1
- j_l(k_1R_\text{c})   k_1 \Psi'_l(k_2R_\text{c})   Z_2))   Z_3 + 
a_{lm}^\text{0,II,A} (j_l(k_2R_\text{c})   k_2 \Psi'_l(k_1R_\text{c})   Z_1\\ &- j_l(k_1R_\text{c})   k_1 \Psi'_l(k_2R_\text{c})   Z_2)  (h^+_l(k_3R_\text{s})   k_3 \Psi'_l(k_2R_\text{s})   Z_2  
-j_l(k_2R_\text{s})   k_2 \xi'_l(k_3R_\text{s})   Z_3))\\
&/(h^+_l(k_3R_\text{s})   k_3   Z_2  (-h^+_l(k_2R_\text{c})   k_2 \Psi'_l(k_1R_\text{c}) \Psi'_l(k_2R_\text{s})   Z_1 
+ j_l(k_2R_\text{c})   k_2 \Psi'_l(k_1R_\text{c}) \xi'_l(k_2R_\text{s})  Z_1 + \\
& j_l(k_1R_\text{c})   k_1  (\psi_(k_2R_\text{s}) \xi'_l(k_2R_\text{c}) 
-\Psi'_l(k_2R_\text{c}) \xi'_l(k_2R_\text{s}))   Z_2) + 
k_2 \xi'_l(k_3R_\text{s})  (-h^+_l(k_2R_\text{s})  j_l(k_2R_\text{c})   k_2 \Psi'_l(k_1R_\text{c})   Z_1 \\
&+ h^+_l(k_2R_\text{c})  j_l(k_2R_\text{s}) k_2 \Psi'_l(k_1R_\text{c})   Z_1 + h^+_l(k_2R_\text{s})  j_l(k_1R_\text{c})   k_1 \Psi'_l(k_2R_\text{c})   Z_2
- j_l(k_1R_\text{c})  j_l(k_2R_\text{s})   k_1 \xi'_l(k_2R_\text{c})   Z_2)   Z_3) ,
\end{split}
\end{equation}
\begin{equation}
\label{eq:almIIB_pen_core-shell}
\begin{split}
a_{lm}^\text{II,B} = &
(-a_{lm}^\text{0,III}  k_2  (h^+_l(k_3R_\text{s}) \Psi'_l(k_3R_\text{s})
- j_l(k_3R_\text{s}) \xi'_l(k_3R_\text{s}))   Z_2  (h^+_l(k_2R_\text{c}) 
k_2 \Psi'_l(k_1R_\text{c})   Z_1
- j_l(k_1R_\text{c})   k_1 \xi'_l(k_2R_\text{c})   Z_2) + \\
& ((a_{lm}^\text{0,II,B} h^+_l(k_2R_\text{c}) - a_{lm}^\text{0,II,A} j_l(k_2R_\text{c}))   k_2 \Psi'_l(k_1R_\text{c})   Z_1
+ a_{lm}^\text{0,I}  k_2  (h^+_l(k_1R_\text{c}) \Psi'_l(k_1R_\text{c}) - j_l(k_1R_\text{c}) \xi'_l(k_1R_\text{c}))   Z_2  \\
&+j_l(k_1R_\text{c})   k_1  (a_{lm}^\text{0,II,A}\Psi'_l(k_2R_\text{c})
- a_{lm}^\text{0,II,B}\xi'_l(k_2R_\text{c}))   Z_2)  (h^+_l(k_3R_\text{s}) k_3 \xi'_l(k_2R_\text{s})   Z_2
- h^+_l(k_2R_\text{s})   k_2 \xi'_l(k_3R_\text{s})   Z_3)) \\
&    /(h^+_l(k_3R_\text{s})   k_3   Z_2  (-h^+_l(k_2R_\text{c})   k_2 \Psi'_l(k_1R_\text{c}) \Psi'_l(k_2R_\text{s})   Z_1
+ j_l(k_2R_\text{c})   k_2 \Psi'_l(k_1R_\text{c}) \xi'_l(k_2R_\text{s}) Z_1 + \\
& j_l(k_1R_\text{c})   k_1  (\Psi'_l(k_2R_\text{s}) \xi'_l(k_2R_\text{c}) 
-\Psi'_l(k_2R_\text{c}) \xi'_l(k_2R_\text{s}))   Z_2) 
+ k_2 \xi'_l(k_3R_\text{s})  (-h^+_l(k_2R_\text{s})  j_l(k_2R_\text{c})  k_2 \Psi'_l(k_1R_\text{c})   Z_1 \\
&+ h^+_l(k_2R_\text{c})  j_l(k_2R_\text{s})   k_2 \Psi'_l(k_1R_\text{c})   Z_1+ h^+_l(k_2R_\text{s})  j_l(k_1R_\text{c})   k_1 \Psi'_l(k_2R_\text{c})  Z_2
- j_l(k_1R_\text{c})  j_l(k_2R_\text{s})   k_1 \xi'_l(k_2R_\text{c})   Z_2)   Z_3) ,
\end{split}
\end{equation}
\begin{equation}
\label{eq:almIII_pen_core-shell}
\begin{split}
a_{lm}^\text{III} = &
(a_{lm}^\text{0,III} j_l(k_3R_\text{s})   k_3   Z_2  (-h^+_l(k_2R_\text{c})   k_2 \Psi'_l(k_1R_\text{c}) \Psi'_l(k_2R_\text{s}) Z_1
+ \\
& j_l(k_2R_\text{c})   k_2 \Psi'_l(k_1R_\text{c}) \xi'_l(k_2R_\text{s}) Z_1 + 
j_l(k_1R_\text{c})   k_1  (\psi_(k_2R_\text{s}) \xi'_l(k_2R_\text{c})  
-\Psi'_l(k_2R_\text{c}) \xi'_l(k_2R_\text{s}))   Z_2) + \\
& a_{lm}^\text{0,III}  k_2 \Psi'_l(k_3R_\text{s})  (-h^+_l(k_2R_\text{s})  j_l(k_2R_\text{c})   k_2 \Psi'_l(k_1R_\text{c})   Z_1
+ h^+_l(k_2R_\text{c})  j_l(k_2R_\text{s})   k_2 \Psi'_l(k_1R_\text{c})   Z_1 + \\
& h^+_l(k_2R_\text{s})  j_l(k_1R_\text{c})   k_1 \Psi'_l(k_2R_\text{c})Z_2
- j_l(k_1R_\text{c})  j_l(k_2R_\text{s})   k_1 \xi'_l(k_2R_\text{c})   Z_2)   Z_3 +
k_3  (h^+_l(k_2R_\text{s}) \Psi'_l(k_2R_\text{s})\\
&- j_l(k_2R_\text{s}) \xi'_l(k_2R_\text{s}))  ((a_{lm}^\text{0,II,B} h^+_l(k_2R_\text{c}) - a_{lm}^\text{0,II,A} j_l(k_2R_\text{c}))   k_2 \Psi'_l(k_1R_\text{c})   Z_1 
+ a_{lm}^\text{0,I}  k_2  (h^+_l(k_1R_\text{c}) \Psi'_l(k_1R_\text{c}) - \\
& j_l(k_1R_\text{c}) \xi'_l(k_1R_\text{c}))   Z_2 + 
j_l(k_1R_\text{c})   k_1  (a_{lm}^\text{0,II,A}\Psi'_l(k_2R_\text{c})
- a_{lm}^\text{0,II,B}\xi'_l(k_2R_\text{c}))   Z_2)   Z_3) \\
& /(h^+_l(k_3R_\text{s})   k_3   Z_2  (h^+_l(k_2R_\text{c})   k_2 \Psi'_l(k_1R_\text{c}) \Psi'_l(k_2R_\text{s})   Z_1
- j_l(k_2R_\text{c})   k_2 \Psi'_l(k_1R_\text{c}) \xi'_l(k_2R_\text{s})   Z_1 + \\
& j_l(k_1R_\text{c})   k_1  (-\Psi'_l(k_2R_\text{s}) \xi'_l(k_2R_\text{c})
+\Psi'_l(k_2R_\text{c}) \xi'_l(k_2R_\text{s}))   Z_2) + 
k_2 \xi'_l(k_3R_\text{s})  (h^+_l(k_2R_\text{s})  j_l(k_2R_\text{c})   k_2 \Psi'_l(k_1R_\text{c})   Z_1\\
& - h^+_l(k_2R_\text{c})  j_l(k_2R_\text{s})   k_2 \Psi'_l(k_1R_\text{c})   Z_1 - h^+_l(k_2R_\text{s})  j_l(k_1R_\text{c})  k_1 \Psi'_l(k_2R_\text{c})   Z_2 
+j_l(k_1R_\text{c})  j_l(k_2R_\text{s})   k_1 \xi'_l(k_2R_\text{c})   Z_2)   Z_3) ,
\end{split}
\end{equation}
\begin{equation}
\label{eq:blmI_pen_core-shell}
\begin{split}
b_{lm}^\text{I} = &
((h^+_l(k_3R_\text{s})  k_3   Z_3  (b_{lm}^\text{0,II,B}   Z_1  k_1   (-h^+_l(k_2R_\text{c}) \Psi'_l(k_2R_\text{c}) + 
j_l(k_2R_\text{c}) \xi'_l(k_2R_\text{c})) \xi'_l(k_2R_\text{s}) \\
&+ b_{lm}^\text{0,I}  k_2  Z_2 \xi'_l(k_1R_\text{c})  (-h^+_l(k_2R_\text{c}) \Psi'_l(k_2R_\text{s}) + 
j_l(k_2R_\text{c}) \xi'_l(k_2R_\text{s}))) 
+ b_{lm}^\text{0,III}  k_1  Z_1  Z_2   k_2  (h^+_l(k_2R_\text{c}) \Psi'_l(k_2R_\text{c}) -\\
& j_l(k_2R_\text{c}) \xi'_l(k_2R_\text{c}))  (h^+_l(k_3R_\text{s}) \Psi'_l(k_3R_\text{s})
- j_l(k_3R_\text{s}) \xi'_l(k_3R_\text{s})) + 
b_{lm}^\text{0,I} h^+_l(k_1R_\text{c})  h^+_l(k_3R_\text{s})   Z_1  k_1   Z_3  k_3  (psi_(k_2R_\text{s}) \xi'_l(k_2R_\text{c}) \\
&-\Psi'_l(k_2R_\text{c}) \xi'_l(k_2R_\text{s}))) + 
k_2  Z_2 \xi'_l(k_3R_\text{s})  (b_{lm}^\text{0,I} (-h^+_l(k_2R_\text{s})  j_l(k_2R_\text{c}) 
+ h^+_l(k_2R_\text{c})  j_l(k_2R_\text{s}))   k_2   Z_2\xi'_l(k_1R_\text{c}) + \\
& b_{lm}^\text{0,II,B} h^+_l(k_2R_\text{s})  Z_1   k_1  (h^+_l(k_2R_\text{c}) \Psi'_l(k_2R_\text{c})
- j_l(k_2R_\text{c}) \xi'_l(k_2R_\text{c})) + 
b_{lm}^\text{0,I} h^+_l(k_1R_\text{c})  Z_1   k_1  (h^+_l(k_2R_\text{s}) \Psi'_l(k_2R_\text{c})\\
&- j_l(k_2R_\text{s}) \xi'_l(k_2R_\text{c}))) 
+ b_{lm}^\text{0,II,A} Z_1   k_1  (h^+_l(k_2R_\text{c}) \Psi'_l(k_2R_\text{c})
- j_l(k_2R_\text{c}) \xi'_l(k_2R_\text{c})) (h^+_l(k_3R_\text{s})   Z_3  k_3 \Psi'_l(k_2R_\text{s}) - \\
& j_l(k_2R_\text{s})   k_2  Z_2 \xi'_l(k_3R_\text{s}) ))\\ 
&/(h^+_l(k_3R_\text{s})   Z_3  k_3  (h^+_l(k_2R_\text{c})   Z_2  k_2 \Psi'_l(k_1R_\text{c}) \Psi'_l(k_2R_\text{s})
- j_l(k_2R_\text{c})   k_2  Z_2 \Psi'_l(k_1R_\text{c}) \xi'_l(k_2R_\text{s}) + \\
& j_l(k_1R_\text{c})  Z_1   k_1  (-\Psi'_l(k_2R_\text{s}) \xi'_l(k_2R_\text{c})
+\Psi'_l(k_2R_\text{c}) \xi'_l(k_2R_\text{s}))) + 
k_2  Z_2 \xi'_l(k_3R_\text{s})  (h^+_l(k_2R_\text{s})  j_l(k_2R_\text{c})   Z_2  k_2 \Psi'_l(k_1R_\text{c})\\
&- h^+_l(k_2R_\text{c})  j_l(k_2R_\text{s})   Z_2  k_2 \Psi'_l(k_1R_\text{c}) 
- h^+_l(k_2R_\text{s})  j_l(k_1R_\text{c})  Z_1   k_1 \Psi'_l(k_2R_\text{c})
+j_l(k_1R_\text{c})  j_l(k_2R_\text{s})   k_1  Z_1 \xi'_l(k_2R_\text{c}))) ,
\end{split}
\end{equation}
\begin{equation}
\label{eq:blmIIA_pen_core-shell}
\begin{split}
b_{lm}^\text{II,A} = &
((b_{lm}^\text{0,III}  k_2  Z_2  (h^+_l(k_3R_\text{s}) \Psi'_l(k_3R_\text{s}) - j_l(k_3R_\text{s}) \xi'_l(k_3R_\text{s}))  (j_l(k_2R_\text{c})   Z_2  k_2 \Psi'_l(k_1R_\text{c})
- j_l(k_1R_\text{c})  Z_1   k_1 \Psi'_l(k_2R_\text{c})) + \\
& h^+_l(k_3R_\text{s})   k_3  Z_3  (b_{lm}^\text{0,I} Z_2   k_2 \Psi'_l(k_2R_\text{s})
(-h^+_l(k_1R_\text{c}) \Psi'_l(k_1R_\text{c}) 
+ j_l(k_1R_\text{c}) \xi'_l(k_1R_\text{c})) + \\
& b_{lm}^\text{0,II,B}\xi'_l(k_2R_\text{s})
(-j_l(k_2R_\text{c})   Z_2  k_2 \Psi'_l(k_1R_\text{c}) + j_l(k_1R_\text{c})   Z_1  k_1 \Psi'_l(k_2R_\text{c}))))\\
&+ k_2  Z_2 \xi'_l(k_3R_\text{s})  (b_{lm}^\text{0,I} j_l(k_2R_\text{s})  Z_2   k_2  (h^+_l(k_1R_\text{c}) \Psi'_l(k_1R_\text{c}) - 
j_l(k_1R_\text{c}) \xi'_l(k_1R_\text{c})) +\\
& b_{lm}^\text{0,II,B} h^+_l(k_2R_\text{s})  (j_l(k_2R_\text{c})  Z_2   k_2 \Psi'_l(k_1R_\text{c}) - 
j_l(k_1R_\text{c})  Z_1   k_1 \Psi'_l(k_2R_\text{c}))) \\
& + b_{lm}^\text{0,II,A} (j_l(k_2R_\text{c})   Z_2  k_2 \Psi'_l(k_1R_\text{c}) - j_l(k_1R_\text{c})  Z_1   k_1 \Psi'_l(k_2R_\text{c})) 
(h^+_l(k_3R_\text{s})  Z_3   k_3 \Psi'_l(k_2R_\text{s}) 
- j_l(k_2R_\text{s})   k_2  Z_2 \xi'_l(k_3R_\text{s})))\\
&/(h^+_l(k_3R_\text{s})  Z_3   k_3  (-h^+_l(k_2R_\text{c})   k_2  Z_2 \Psi'_l(k_1R_\text{c}) \Psi'_l(k_2R_\text{s}) 
+ j_l(k_2R_\text{c})   k_2  Z_2 \Psi'_l(k_1R_\text{c}) \xi'_l(k_2R_\text{s}) + \\
& j_l(k_1R_\text{c})   Z_1  k_1  (psi_(k_2R_\text{s}) \xi'_l(k_2R_\text{c})
-\Psi'_l(k_2R_\text{c}) \xi'_l(k_2R_\text{s}))) + 
k_2  Z_2 \xi'_l(k_3R_\text{s})  (-h^+_l(k_2R_\text{s})  j_l(k_2R_\text{c})  Z_2   k_2 \Psi'_l(k_1R_\text{c}) \\
&+ h^+_l(k_2R_\text{c})  j_l(k_2R_\text{s})  Z_2   k_2 \Psi'_l(k_1R_\text{c}) + 
h^+_l(k_2R_\text{s})  j_l(k_1R_\text{c})   Z_1  k_1 \Psi'_l(k_2R_\text{c})
- j_l(k_1R_\text{c})  j_l(k_2R_\text{s})  Z_1   k_1 \xi'_l(k_2R_\text{c}))) ,
\end{split}
\end{equation}
\begin{equation}
\label{eq:blmIIB_pen_core-shell}
\begin{split}
b_{lm}^\text{II,B} = &
(-b_{lm}^\text{0,III}  k_2  Z_2  (h^+_l(k_3R_\text{s}) \Psi'_l(k_3R_\text{s}) - j_l(k_3R_\text{s}) \xi'_l(k_3R_\text{s}))  (h^+_l(k_2R_\text{c})  Z_2   k_2 \Psi'_l(k_1R_\text{c}) \\
&- j_l(k_1R_\text{c})   k_1  Z_1 \xi'_l(k_2R_\text{c}) ) + ((b_{lm}^\text{0,II,B} h^+_l(k_2R_\text{c}) - b_{lm}^\text{0,II,A} j_l(k_2R_\text{c}))   Z_2  k_2 \Psi'_l(k_1R_\text{c})
+ b_{lm}^\text{0,I} Z_2   k_2  (h^+_l(k_1R_\text{c}) \Psi'_l(k_1R_\text{c}) - \\
& j_l(k_1R_\text{c}) \xi'_l(k_1R_\text{c})) 
+  j_l(k_1R_\text{c})   k_1  Z_1  (b_{lm}^\text{0,II,A}\Psi'_l(k_2R_\text{c}) 
- b_{lm}^\text{0,II,B}\xi'_l(k_2R_\text{c})))  (h^+_l(k_3R_\text{s})  Z_3   k_3 \xi'_l(k_2R_\text{s})\\
&-  h^+_l(k_2R_\text{s})   k_2  Z_2 \xi'_l(k_3R_\text{s})))\\
& /(h^+_l(k_3R_\text{s})   k_3   Z_3  (-h^+_l(k_2R_\text{c})   k_2  Z_2 \Psi'_l(k_1R_\text{c}) \Psi'_l(k_2R_\text{s})
+  j_l(k_2R_\text{c})   k_2  Z_2 \Psi'_l(k_1R_\text{c}) \xi'_l(k_2R_\text{s}) + \\
& j_l(k_1R_\text{c})  Z_1   k_1  (\psi(k_2R_\text{s}) \xi'_l(k_2R_\text{c}) 
-\Psi'_l(k_2R_\text{c}) \xi'_l(k_2R_\text{s}))) + 
k_2  Z_2 \xi'_l(k_3R_\text{s})  (-h^+_l(k_2R_\text{s})  j_l(k_2R_\text{c})   k_2  Z_2 \Psi'_l(k_1R_\text{c}) \\
&+ h^+_l(k_2R_\text{c})  j_l(k_2R_\text{s})   k_2  Z_2 \Psi'_l(k_1R_\text{c}) + 
h^+_l(k_2R_\text{s})  j_l(k_1R_\text{c})   Z_1  k_1 \Psi'_l(k_2R_\text{c})
- j_l(k_1R_\text{c})  j_l(k_2R_\text{s})  Z_1   k_1 \xi'_l(k_2R_\text{c}))) ,
\end{split}
\end{equation}
\begin{equation}
\label{eq:blmIII_pen_core-shell}
\begin{split}
b_{lm}^\text{III} = &
(b_{lm}^\text{0,III} j_l(k_3R_\text{s})   k_3   Z_3  (-h^+_l(k_2R_\text{c})   Z_2  k_2 \Psi'_l(k_1R_\text{c}) \Psi'_l(k_2R_\text{s})
+ j_l(k_2R_\text{c})  Z_2   k_2 \Psi'_l(k_1R_\text{c}) \xi'_l(k_2R_\text{s}) +\\
& j_l(k_1R_\text{c})   k_1  Z_1  (\psi(k_2R_\text{s}) \xi'_l(k_2R_\text{c})
-\Psi'_l(k_2R_\text{c}) \xi'_l(k_2R_\text{s})))
+ b_{lm}^\text{0,III}  Z_2  k_2 \Psi'_l(k_3R_\text{s})  (-h^+_l(k_2R_\text{s})  j_l(k_2R_\text{c})  Z_2   k_2 \Psi'_l(k_1R_\text{c}) \\
& + h^+_l(k_2R_\text{c})  j_l(k_2R_\text{s})  Z_2   k_2 \Psi'_l(k_1R_\text{c}) +
h^+_l(k_2R_\text{s})  j_l(k_1R_\text{c})   k_1  Z_1 \Psi'_l(k_2R_\text{c})
- j_l(k_1R_\text{c})  j_l(k_2R_\text{s})   k_1  Z_1 \xi'_l(k_2R_\text{c})) + \\
& k_3  Z_3  (h^+_l(k_2R_\text{s}) \Psi'_l(k_2R_\text{s})
- j_l(k_2R_\text{s}) \xi'_l(k_2R_\text{s}))  ((b_{lm}^\text{0,II,B} h^+_l(k_2R_\text{c}) - b_{lm}^\text{0,II,A} j_l(k_2R_\text{c}))  Z_2   k_2 \Psi'_l(k_1R_\text{c})\\
&+ b_{lm}^\text{0,I} Z_2   k_2  (h^+_l(k_1R_\text{c}) \Psi'_l(k_1R_\text{c}) - j_l(k_1R_\text{c}) \xi'_l(k_1R_\text{c})) + 
j_l(k_1R_\text{c})   k_1  Z_1
\times(b_{lm}^\text{0,II,A}\Psi'_l(k_2R_\text{c})- b_{lm}^\text{0,II,B}\xi'_l(k_2R_\text{c}))))\\
&/(h^+_l(k_3R_\text{s})  Z_3   k_3  (h^+_l(k_2R_\text{c})  Z_2   k_2 \Psi'_l(k_1R_\text{c}) \Psi'_l(k_2R_\text{s})
- j_l(k_2R_\text{c})   k_2  Z_2 \Psi'_l(k_1R_\text{c}) \xi'_l(k_2R_\text{s}) + \\
& j_l(k_1R_\text{c})   Z_1  k_1  (-\Psi'_l(k_2R_\text{s}) \xi'_l(k_2R_\text{c})
+\Psi'_l(k_2R_\text{c}) \xi'_l(k_2R_\text{s}))) + 
k_2  Z_2 \xi'_l(k_3R_\text{s})  (h^+_l(k_2R_\text{s})  j_l(k_2R_\text{c})   Z_2  k_2 \Psi'_l(k_1R_\text{c}) \\
&- h^+_l(k_2R_\text{c})  j_l(k_2R_\text{s})   Z_2  k_2 \Psi'_l(k_1R_\text{c}) - h^+_l(k_2R_\text{s})  j_l(k_1R_\text{c})   Z_1  k_1 \Psi'_l(k_2R_\text{c})
+j_l(k_1R_\text{c})  j_l(k_2R_\text{s})  Z_1   k_1 \xi'_l(k_2R_\text{c}))) .
\end{split}
\end{equation}

\subsection{EEL probability for core--shell penetrating trajectories}\label{App:EEL}

The total EEL probability for an electron beam traversing both the core and shell of a core--shell NP is given by
\begin{align}
\Gamma_\text{EEL}(\omega) =
\Gamma_\text{bulk}^\text{I}(\omega) + \Gamma_\text{bulk}^\text{II}(\omega) + \Gamma_\text{begr}^\text{I}(\omega) +
\Gamma_\text{begr}^\text{II,A}(\omega) + \Gamma_\text{begr}^\text{II,B}(\omega) + \Gamma_\text{surf}(\omega) \ .
\end{align}
The individual terms read 
\begin{align}
\label{eq:GbulkI_pen_core}
\Gamma_{\text{bulk}}^\text{I}(\omega) &=   -\frac{e^2z_\text{c}}{2\pi^2\hbar\varepsilon_0v^2}\text{Im} \left[  \frac{1}{\gamma_1^2\varepsilon_1}\ln\left( \left( \frac{q_\text{c}\gamma_1v}{\omega} \right)^2 +1 \right) \right] ,
\end{align}
\begin{align}
\label{eq:GbulkII_pen_core}
\Gamma_{\text{bulk}}^\text{II}(\omega)  =  -\frac{e^2(z_\text{s}-z_\text{c})}{2\pi^2\hbar\varepsilon_0v^2}\text{Im} \left[\frac{1}{\gamma_2^2\varepsilon_2}\ln\left( \left( \frac{q_\text{c}\gamma_2v}{\omega} \right)^2 +1 \right) \right],
\end{align}
\begin{align}
\label{eq:GbegrI_pen_corr}
\Gamma_\text{begr}^\text{I}(\omega) = \frac{e}{\pi \hbar\omega} \text{Re} & \Bigg[ \sum_{l,m}  \int_{-z_\text{c}}^{z_\text{c}}\text{d}z \ \frac{\text{e}^{\mathrm{i} m \phi_0}\text{e}^{-\mathrm{i} \omega z/v}}{\sqrt{l(l+1)}} 
\times \bigg(mb^\text{I}_{lm}j_l(k_1r)Y_l^m(\theta,0) - \frac{a^\text{I}_{lm}}{k_1b}(\mathcal{J}^+_{lm}(k_1z) + \mathcal{J}^-_{lm}(k_1z)) 
\nonumber\\
& - mb^\text{0,II,A}_{lm}j_l(k_2r)Y_l^m(\theta,0) + \frac{a^\text{0,II,A}_{lm}}{k_2b}(\mathcal{J}^+_{lm}(k_2z) + \mathcal{J}^-_{lm}(k_2z))\bigg) \Bigg] ,
\end{align}
\begin{align}
\label{GbegrIIA_pen_core}
\Gamma_\text{begr}^\text{II,A}(\omega) &= \frac{e}{\pi \hbar \omega} \text{Re} \Bigg[ 
\sum_{l,m} \int_{-z_\text{s}}^{z_\text{s}} \text{d}z \ \frac{\text{e}^{\mathrm{i} m\phi_0}\text{e}^{-\mathrm{i}\omega z/v}}{\sqrt{l(l+1)}}\Big[ m b_{lm}^\text{II,A}h^+_l(k_2r)Y_l^m(\theta,0)
-\frac{a_{lm}^\text{II,A}}{k_2b}(\mathcal{H}^+_{lm}(k_2z) + 
\mathcal{H}_{lm}^-(k_2z)) \Big] 
\nonumber\\
&- \int_{-z_\text{c}}^{z_\text{c}} \text{d}z \ \frac{\text{e}^{\mathrm{i} m\phi_0}\text{e}^{-\mathrm{i}\omega z/v}}{\sqrt{l(l+1)}}\Big[ m b_{lm}^\text{II,A}h^+_l(k_2r)Y_l^m(\theta,0)
-\frac{a_{lm}^\text{II,A}}{k_2b}(\mathcal{H}^+_{lm}(k_2z) + \mathcal{H}_{lm}^-(k_2z)) \Big] \Bigg],
\end{align}
\begin{align}
\label{eq:GbegrIIB_pen_corr}
\Gamma_\text{begr}^\text{II,B}(\omega) &= 
\frac{e}{\pi \hbar \omega} \text{Re} \Bigg[ 
\sum_{l,m} \int_{-z_\text{s}}^{z_\text{s}} \text{d}z \ \frac{\text{e}^{\mathrm{i} m\phi_0}\text{e}^{-\mathrm{i}\omega z/v}}{\sqrt{l(l+1)}} \times \bigg( m b_{lm}^\text{II,B}j_l(k_2r)Y_l^m(\theta,0) -\frac{a_{lm}^\text{II,B}}{k_2b}(\mathcal{J}^+_{lm}(k_2z) + \mathcal{J}_{lm}^-(k_2z)) 
\nonumber \\
& - mb_{lm}^\text{0,III}j_l(k_3r)Y_l^m(\theta,0) -\frac{a_{lm}^\text{0,III}}{k_3r}(\mathcal{J}^+_{lm}(k_3z) + \mathcal{J}_{lm}^-(k_3z)) \bigg)
- \int_{-z_\text{c}}^{z_\text{c}} \text{d}z \ \frac{\text{e}^{\mathrm{i} m\phi_0}\text{e}^{-\mathrm{i}\omega z/v}}{\sqrt{l(l+1)}}\Big[ m b_{lm}^\text{II,B}j_l(k_2r)Y_l^m(\theta,0) 
\nonumber\\
&-\frac{a_{lm}^\text{II,B}}{k_2b}(\mathcal{J}^+_{lm}(k_2z) + \mathcal{J}_{lm}^-(k_2z)) \Big] \Bigg] .
\end{align}
\begin{align}
\label{eq:Gsurf_pen_core}
\Gamma_\text{surf}(\omega) 
&= \frac{e}{\pi \hbar \omega} \text{Re} \Bigg[ 
\sum_{l,m} \text{e}^{\mathrm{i} m\phi_0}\Bigg(\frac{K_m\left(\frac{\omega b}{v\gamma_3} \right)}{\mathrm{i} k_3\sqrt{l(l+1)}}\left( m b_{lm}^\text{III}\mathcal{M}_{lm}^{+*} 
-a_{lm}^\text{III}\frac{\mathcal{N}_{lm}^{+*} }{\beta \gamma_3} \right) 
\nonumber \\
&- \int_{-z_\text{s}}^{z_\text{s}} \text{d}z \ \frac{e^{-\mathrm{i}\omega z/v}}{\sqrt{l(l+1)}}\Big[ m b_{lm}^\text{III}h^+_l(k_3r)Y_l^m(\theta,0) 
-\frac{a_{lm}^\text{III}}{k_3b}(\mathcal{H}^+_{lm}(k_3z) + \mathcal{H}_{lm}^-(k_3z)) \Big] \Bigg) \Bigg] .
\end{align}

\subsection{EEL probability of shell-penetrating trajectories}\label{app:EEL_shell_pen}

To account for the different electron beam trajectories compared to the full core--shell penetrating case, the expansion coefficients of the incident field within the shell $a/b_{lm}^\text{0,II,A}$ and $a/b_{lm}^\text{0,II,B}$ must be modified as
\begin{align}
\label{eq:alm0IIA_pen_shell}
a_{lm}^\text{0,II,A} = \frac{\mathrm{i} k_2^2e}{\varepsilon_2\varepsilon_0\omega} \frac{e^{-\mathrm{i} m\phi_0}}{\sqrt{l(l+1)}} \frac{\mathrm{i}}{b}
\int_{-z_\text{s}}^{z_\text{s}} \text{d}z \ e^{\mathrm{i}\omega z/v}(\mathcal{H}_{lm}^+(k_2z)+\mathcal{H}_{lm}^-(k_2z)),
\end{align}
\begin{align}
\label{eq:blm0IIA_pen_shell}
b_{lm}^\text{0,II,A} =  -\frac{\mathrm{i} k_2^2e}{\varepsilon_2\varepsilon_0\omega} \frac{me^{-\mathrm{i} m\phi_0}}{\sqrt{l(l+1)}} \mathrm{i} k_2 
\int_{-z_\text{s}}^{z_\text{s}} \text{d}z \ e^{\mathrm{i}\omega z/v}h^+_l(k_2z)Y_l^m(\theta,0),
\end{align}
\begin{align}
\label{eq:alm0IIB_pen_shell}
a_{lm}^\text{0,II,B} =  
\frac{\mathrm{i} k_2^2e}{\varepsilon_2\varepsilon_0\omega} \frac{e^{-\mathrm{i} m\phi_0}}{\sqrt{l(l+1)}} \frac{\mathrm{i}}{b} 
\int_{-z_\text{s}}^{z_\text{s}} \text{d}z \ e^{\mathrm{i}\omega z/v}(\mathcal{J}_{lm}^+(k_2z)+\mathcal{J}_{lm}^-(k_2z)),
\end{align}
\begin{align}
\label{eq:blm0IIB_pen_shell}
b_{lm}^\text{0,II,B} =  -\frac{\mathrm{i} k_2^2e}{\varepsilon_2\varepsilon_0\omega} \frac{me^{-\mathrm{i} m\phi_0}}{\sqrt{l(l+1)}} \mathrm{i} k_2 \int_{-z_\text{s}}^{z_\text{s}} \text{d}z \ e^{\mathrm{i}\omega z/v}j_l(k_2z)Y_l^m(\theta,0) .
\end{align}
The total EEL probability for an electron beam penetrating the shell of a core--shell NP is given by
\begin{align}\label{eq:core_shell_decomp}
\Gamma_\text{EEL}(\omega) = \Gamma^\text{II}_\text{bulk}(\omega) + \Gamma^\text{II,A}_\text{begr}(\omega) + \Gamma^\text{II,B}_\text{begr}(\omega) + \Gamma_\text{surf}(\omega) ,
\end{align}
with
\begin{align}
\label{eq:GbulkII_pen_shell}
\Gamma_{\text{bulk}}^\text{II}(\omega)  =  -\frac{e^2z_\text{s}}{2\pi^2\hbar\varepsilon_0v^2}\text{Im} \left[ \frac{1}{\gamma_2^2\varepsilon_2}\ln\left( \left( \frac{q_\text{c}\gamma_2v}{\omega} \right)^2 +1 \right) \right] ,
\end{align}
\begin{align}
\label{GbegrIIA_pen_shell}
\Gamma_\text{begr}^\text{II,A}(\omega) &= \frac{e}{\pi \hbar \omega} \text{Re} 
\Bigg[ 
\sum_{l,m} \int_{-z_\text{s}}^{z_\text{s}} \text{d}z \ \frac{\text{e}^{\mathrm{i} m\phi_0}\text{e}^{-\mathrm{i}\omega z/v}}{\sqrt{l(l+1)}} \times \bigg( m b_{lm}^\text{II,A}h^+_l(k_2r)Y_l^m(\theta,0) 
    -\frac{a_{lm}^\text{II,A}}{k_2b}(\mathcal{H}^+_{lm}(k_2z) + \mathcal{H}_{lm}^-(k_2z)) \bigg) \Bigg],
\end{align}
\begin{align}
\label{eq:GbegrIIB_pen_corr_shell}
\Gamma_\text{begr}^\text{II,B}(\omega) &= \frac{e}{\pi \hbar \omega} \text{Re} 
\Bigg[ 
\sum_{l,m} \int_{-z_\text{s}}^{z_\text{s}} \text{d}z \ \frac{\text{e}^{\mathrm{i} m\phi_0}\text{e}^{-\mathrm{i}\omega z/v}}{\sqrt{l(l+1)}}
\times \bigg( m b_{lm}^\text{II,B}j_l(k_2r)Y_l^m(\theta,0) -\frac{a_{lm}^\text{II,B}}{k_2b}(\mathcal{J}^+_{lm}(k_2z) + \mathcal{J}_{lm}^-(k_2z)) \\
& - mb_{lm}^\text{0,III}j_l(k_3r)Y_l^m(\theta,0) -\frac{a_{lm}^\text{0,III}}{k_3r}(\mathcal{J}^+_{lm}(k_3z) + \mathcal{J}_{lm}^-(k_3z)) \bigg) \Bigg] ,
\end{align}
while the surface contribution $\Gamma_\text{surf}(\omega)$ is identical to the fully penetrating case and is therefore still given by Eq.~\eqref{eq:Gsurf_pen_core}. The correction accounting for the direct electron field has already been incorporated into $\Gamma_\text{begr}^\text{II,B}(\omega)$.
\end{widetext}

\bibliography{references_2}

\end{document}